%% file: article.tex
\documentclass{elsart}

\usepackage{natbib}
\usepackage{amsmath,amsfonts}
\usepackage{epsfig}
\usepackage{enumerate}
\usepackage{feynmp}
\usepackage{subfigure}

\newcommand{\comm}[2]  		{\big[  #1 \, , \, #2 \big] }
\newcommand{\mean}[1]		{\langle #1 \rangle}
\newcommand{\bra}[1]   		{\langle #1 |}
\newcommand{\ket}[1]   		{| #1 \rangle}
\newcommand{\bracket}[2]	{\langle #1 | #2 \rangle}
\newcommand{\dt}			{\mathrm{d}t}
\newcommand{\dz}			{\mathrm{d}z}
\newcommand{\dtau}			{\mathrm{d}\tau}
\newcommand{\Tr}			{\mathrm{Tr}}
\newcommand{\openone}		{\leavevmode\hbox{\small1\normalsize\kern-.33em1}}
\newcommand{\eps}			{\varepsilon}
\newcommand{\ppf}[2]		{\mathcal{P}\!\!\left(\frac{#1}{#2}\right)}
\newcommand{\pint}			{\mathcal{P}\!\!\!\!\!\!\int}

\begin{document}

%---------------------------------------------------------------------------

\begin{frontmatter}
\title{Generalized {N}ozi\`eres--{D}e~{D}ominicis approach to transport through nano--junctions}
\author{B. Braunecker and P.-A. Bar\`es}
\address{Institut de Physique Th\'eorique, D\'epartement de Physique,
\'Ecole Polytechnique F\'ed\'erale de Lausanne EPFL, CH--1015 Lausanne, 
Switzerland}

\begin{abstract}
We investigate the transport properties of a model of an interacting electronic
resonant level system. The hybridization of the localized resonant level is treated in
perturbation and the contributions to all orders are computed. This includes
an exact treatment of the electron scattering from the localized level. The renormalization
of the direct tunneling between the electrodes arising from the Coulomb repulsion of 
trapped particles is taken into account. Further, we provide a detailed solution
of the system of coupled singular integral equations which determine the propagator
in the interacting system.
\end{abstract}

\end{frontmatter}

%-----------------------------------------------------------------------------------------------------
\begin{fmffile}{fmfdiags}     %for Feynman diagrams

%-----------------------------------------------------------------------------------------------------
\section{Introduction}

Over 30 years ago, Nozi\`eres and De~Dominicis (ND) \cite{ND69} proposed an ingenious 
solution of the x-ray absorption/emission edge singularity problem in
metals. Their solution, being non-perturbative, allowed to circumvent
the difficulties arising from the Anderson orthogonality catastrophe \cite{Anderson67}, 
and revealed the singular behavior of the absorption/emission
resonance. Soon thereafter, Yuval and Anderson \cite{YA70} used the formal
resemblance between the x-ray problem and the Kondo spin $1/2$ model to extend the
method to describe the vacuum-to-vacuum amplitude (or the partition function in the
finite temperature case). They showed that the latter is formally equivalent
to the partition function of a classical one-dimensional gas of particles with logarithmic
interactions. In a subsequent work, Anderson, Yuval and Hamann \cite{AYH70}
introduced renormalization group methods to solve this partition function.
Another formal correspondence to the x-ray problem can be found in the description
of the tunneling of an electron through a localized state in a junction.
For such a system, Matveev and Larkin \cite{Matveev92} applied ND's technique
to express the resonant behavior of the tunneling current.

The common property shared among these examples is a time-dependent interaction between 
conduction electrons in certain quantum channels with a localized structureless 
scatterer (a hole, an impurity spin, or a trapped electron). Since the interaction does
not change the internal state of the scatterer, a reduction of the many-body problem
into one-body problems is possible. The latter can then be solved by the
methods devised by ND.

In the examples above it was assumed that electrons in different quantum
channels (such as angular momentum, spin, pseudo spin, or indexing spatially separated electrodes)
do not interact. Generally this results from symmetry arguments and a partial
diagonalization of the Hamiltonian. In the present paper, however, we focus on the
case in which such a decoupling is not possible, and in which particles can be
scattered onto different channels due to the interaction with the localized scatterer.
We choose as an example the physical system considered by Matveev and Larkin
of two electrodes connected to a barrier containing one localized state. However, in contrast
to Matveev and Larkin, we assume a direct coupling of the conduction electrons in 
both electrodes. The method
is illustrated through the computation of the tunneling current, expressed in the form of 
an infinite series (see Eqs.~\eqref{eq:disc:F_W} and \eqref{eq:disc:F_V}
which are the results of this paper), although a large part of the derivations is performed
in the most general case accessible by this approach.

This is a technical paper and we focus mainly on the formal aspects of the problem. 
We expect to come back in the future to physical applications of the generic
problem solved below. 
On the other hand, the physical nature of the system considered below has already been
discussed thoroughly in \cite{Matveev92}. 
For pedagogical reasons details of the computation are written down explicitly.
In particular, we present in the Appendix \ref{sec:int} a full solution of the singular
integral equations determining the one-particle propagators. In this respect, we also 
restrict ourselves to an equilibrium description of the system. 
An extension to the non-equilibrium case will be performed in a subsequent work.

In Sec.~\ref{sec:model} and \ref{sec:dyson} we define the model and derive the Dyson
series for the tunneling current. Sec.~\ref{sec:lowest} and \ref{sec:higher} contain
the central part of the paper in which first ND's method is adapted to the present problem,
and then extended to all orders in the Dyson series following Yuval's and Anderson's ideas.
The application to some simpler problems is discussed in Sec.~\ref{sec:disc}.

%-----------------------------------------------------------------------------------------------------
\section{Model}
\label{sec:model}

The Hamiltonian of the system reads
\begin{equation}
	H = H_0 + H_W + H_V,
\end{equation}
with
\begin{align}
	H_0 &= \sum_{k \alpha} \eps_{k \alpha} \, C^\dagger_{k \alpha} C_{k \alpha} + E_d \, d^\dagger d, \\
	H_W &= \sum_{k k^ \alpha \alpha'} W^{\alpha \alpha'}_{k k'} \, C^\dagger_{k \alpha} C_{k' \alpha'} 
	                                                               d^\dagger d, \\
	H_V &= \sum_{k \alpha} \left( V^\alpha_k \, C_{k \alpha} d^\dagger + \text{h.c.} \right),
\end{align}
where $C^\dagger_{k \alpha}$ and $C_{k \alpha}$ are the creation and annihilation operators of
spinless fermions with momentum $k$ which are characterized by an additional quantum number 
$\alpha = 1, \dots, N$ for some integer $N$.
These particles are assumed to form for every $\alpha$ a Fermi liquid with chemical potential
$\mu_\alpha$, described by $H_0$. The Hamiltonian $H_W$ describes the interaction 
of the particles $(k,\alpha)$ with a localized state, $\phi_d$, which is
characterized by the operators $d^\dagger, d$ and the energy $E_d$.
The hermiticity of $H_W$ imposes that $(W^{\alpha \alpha'}_{k k'})^* = W^{\alpha' \alpha}_{k' k}$.
Finally, $H_V$ denotes the hybridization term with the localized state.

\begin{figure}
\begin{center}
	\input{figure1.tex}
	\caption{Sketch of the physical system underlying the model.}
	\label{fig:system}
\end{center}
\end{figure}
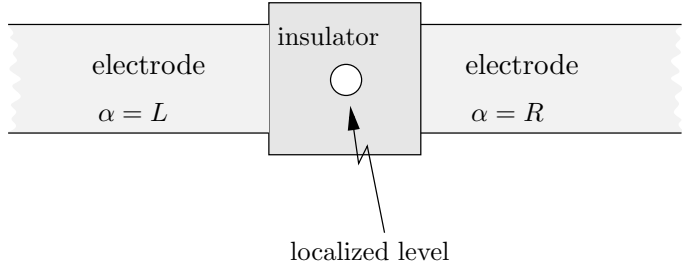

It is essential for the following to assume that this localized state is completely structureless
and that the interaction with the Fermi liquid is short-ranged. Under these assumptions the short-ranged 
potentials can be considered to be separable as
\begin{align}
	W^{\alpha \alpha'}_{k k'} &= W^{\alpha \alpha'} \, (u^\alpha_k)^* u^{\alpha'}_{k'}, \label{eq:model:separable_W}\\
	V^{\alpha}_{k}            &= V^{\alpha} \, u^\alpha_k, \label{eq:model:separable_V}
\end{align}	
with $u^\alpha_k$ some normalized functions that are peaked at the chemical potentials $\mu_\alpha$ where
they take the value $u^\alpha_{k(\mu_\alpha)} = 1$.
It is convenient to define the operators
\begin{equation} \label{eq:model:def_localized_C}
	C_\alpha = \sum_k C_{k \alpha} \, u^\alpha_k \qquad \text{and} \qquad
	C^\dagger_\alpha = \sum_k C^\dagger_{k \alpha} \, (u^\alpha_k)^*
\end{equation}
that annihilate and create a particle localized near $\phi_d$.
The Hamiltonians $H_W$ and $H_V$ become
\begin{align}
	H_W &= \sum_{\alpha \alpha'} W^{\alpha \alpha'} \, C^\dagger_\alpha C_{\alpha'} d^\dagger d, \\
	H_V &= \sum_\alpha \left( V^\alpha \, C_\alpha d^\dagger + \text{h.c.} \right). \label{eq:model:H_V_local}
\end{align}	
As an example for such a system we consider the situation depicted in Fig.~\ref{fig:system}.
An insulating barrier is assumed to be connected on two sides to electrodes containing  
spinless Fermi liquids. The index $\alpha$ labels here the left or right electrode, $\alpha = L,R$.
The localized state $\phi_d$ corresponds to an impurity trap inside the barrier.
Both electrodes are considered as particle reservoirs that are kept at the 
chemical potentials $\mu_L$ and $\mu_R$ (with generally $\mu_L \neq \mu_R$), respectively.
It is assumed that in absence of a particle in the impurity trap the two Fermi
gases are completely independent (we neglect the direct tunneling which can be regarded
as a constant background current).
However, if an electron is trapped at the impurity site, it interacts with the
particles near the junction in the electrodes since the potential created by its charge
is not screened in the insulator. We denote this scattering potential by $W^{\alpha \alpha'}_{k k'}$
where the diagonal terms with respect to $\alpha$, $\alpha = \alpha'$, describe the backscattering
of electrons and the non-diagonal terms, $\alpha \neq \alpha'$, the renormalization of the
background tunneling current.

We intend to calculate the current through the junction, which can be expressed as the time variation
of the number of particles in the electrodes,
\begin{equation}
	J(t) = e \left\langle \frac{\mathrm{d}}{\mathrm{d}t} \left( N^R(t) - N^L(t) \right) \right\rangle,
\end{equation}
where $N^\alpha(t) = \sum_k C_{k \alpha}^\dagger(t) C_{k \alpha}(t)$ is the electron number operator
in the electrode $\alpha$. Its time derivative is determined by the Heisenberg equation
\begin{equation}
	i \frac{\mathrm{d}}{\mathrm{d}t} N^\alpha(t) = \comm{N^\alpha(t)}{H}
	= i \dot{N}^\alpha_W(t) + i \dot{N}^\alpha_V(t),
\end{equation}
where we have defined
\begin{align}
	\dot{N}^\alpha_W(t) 
	&= (-i) \comm{N^\alpha(t)}{H_W(t)} \nonumber \\
	&= (-i) \sum_{\alpha'} 
		\left\{ W^{\alpha \alpha'} C_{\alpha}^\dagger(t) C_{\alpha'}(t) - 
				W^{\alpha' \alpha} C_{\alpha'}^\dagger(t) C_{\alpha}(t) \right\} 
		\, d^\dagger(t) d(t) \\
	\dot{N}^\alpha_V(t) 
	&= (-i) \comm{N^\alpha(t)}{H_V(t)} \nonumber \\
	&= (-i) 
		\left\{ V^{\alpha}     C_{\alpha}^\dagger(t) d(t) - 
				(V^{\alpha})^* C_{\alpha}(t) d^\dagger(t) \right\}.
\end{align}
Therefore the computation of the averages
\begin{align}
	\mean{\dot{N}^\alpha_W(t)} &= 2 \, \mathfrak{Im}\left[ \sum_{\alpha'} W^{\alpha \alpha'}
									 \mean{C^\dagger_\alpha(t) C_{\alpha'}(t) d^\dagger(t) d(t)}\right], \\
	\mean{\dot{N}^\alpha_V(t)} &= 2 \, \mathfrak{Im}\left[ V^{\alpha}
									 \mean{C^\dagger_\alpha(t) d(t)}\right],
\end{align}
which can be expressed by the response functions,
\begin{align}
	F_W^{\alpha \alpha'}(t) &= \mean{C^\dagger_{\alpha}(t) C_{\alpha'}(t) d^\dagger(t) d(t)},\\
	F_V^{\alpha}(t)         &= \mean{C^\dagger_{\alpha}(t) d(t) },
\end{align}
is required.

%-----------------------------------------------------------------------------------------------------

\section{Dyson series}
\label{sec:dyson}

We expand the functions $F_W^{\alpha \alpha'}(t)$ and $F_V^\alpha(t)$ in a zero-temperature 
perturbation series in powers of $H_V$. The interactions due to $H_W$ will be treated in the
exact way indicated by ND. For this purpose, we define an interaction picture with respect to 
$H_1 = H_0 + H_W$ by
\begin{equation}
	i \frac{\mathrm{d}}{\dt} \widehat{A}(t) = \comm{\widehat{A}(t)}{H_1}
\end{equation}
for an arbitrary Schr\"odinger operator $A$.
In order to take averages over operators we impose artificially to the system to be in the same
state $\ket{i}$ at an initial time $t_0$, as well as at the final time $t_1$. We choose
$\ket{i}$ to be an eigenstate of $H_1$ in which the state $\phi_d$ is unoccupied, $d \ket{i} = 0$.
This choice of the average is definitely wrong because the occupancy of $\phi_d$ in the final
state is not well-defined due to the hybridization term $H_V$. 
A more subtle investigation of the problem would involve Keldysh's
formulation of perturbation theory for non-equilibrium processes \cite{Keldysh65} in which 
only the initial, but no final state appears. An extension to the Keldysh formulation 
of the methods described below will be the subject of a future publication. 
In the present paper, however, we choose for pedagogical reasons 
to restrict ourselves to the case in which the system comes back to the state $\ket{i}$ at the
time $t_1$.

The standard perturbation theory tells us that in this case the average value over the operator $A(t)$,
for $t_0 < t < t_1$, can be expressed by the relation
\[
	\mean{A(t)} = \bra{i} T \bigl\{ S(t_0,t_1) \, \widehat{A}(t) \bigr\} \ket{i}
\]
where $T$ denotes the time-order operator, and where
\begin{equation}
	S(t_0,t_1) = \exp \left( -i \int_{t_0}^{t_1} \dtau \, \widehat{H}_V(\tau) \right)
\end{equation}
is the $S$-matrix. In this representation, the functions $F_W^{\alpha \alpha'}(t)$ and $F_V^\alpha(t)$ read
\begin{align*}
	F_W^{\alpha \alpha'}(t) &= \bra{i} T\bigl\{ 
		S(t_0,t_1) \, \widehat{C}^\dagger_{\alpha}(t) \widehat{C}_{\alpha'}(t) 
		              \widehat{d}^\dagger(t) \widehat{d}(t) \bigr\} \ket{i}, \\
	F_V^{\alpha}(t) &= \bra{i} T\bigl\{ 
		S(t_0,t_1) \, \widehat{C}^\dagger_{\alpha}(t) \widehat{d}(t) \bigr\} \ket{i}, 
\end{align*}
or, explicitly,
\begin{align} 
	F_W^{\alpha \alpha'}(t) &= 
	\sum_{n = 0}^\infty \frac{(-i)^{2n}}{(2n)!}
	\int_{t_0}^{t_1} \dtau_1 \dots \dtau_{2n}\\
    &\qquad \qquad
	\bra{i} T\bigl\{ \widehat{H}_V(\tau_1) \dots \widehat{H}_V(\tau_{2n}) \, 
	                 \widehat{C}^\dagger_{\alpha}(t) \widehat{C}_{\alpha'}(t) 
					 \widehat{d}^\dagger(t) \widehat{d}(t) \bigr\} \ket{i}, 
			   																	\label{eq:dyson:Dyson_series_F_W}\\
	F_V^{\alpha}(t) &= 
	\sum_{n = 1}^\infty \frac{(-i)^{2n-1}}{(2n-1)!}
	\int_{t_0}^{t_1} \dtau_1 \dots \dtau_{2n-1}\\
	&\qquad \qquad
    \bra{i} T\bigl\{ \widehat{H}_V(\tau_1) \dots \widehat{H}_V(\tau_{2n-1}) \, 
	                   \widehat{C}^\dagger_{\alpha}(t) \widehat{d}(t) \bigr\} \ket{i}.
					   															\label{eq:dyson:Dyson_series_F_V}
\end{align}

%-----------------------------------------------------------------------------------------------------

\section{Illustration of the method at lowest order}
\label{sec:lowest}

In order to illustrate the method of calculation, let us consider the first non-trivial term $(n=1)$ 
in the expansion \eqref{eq:dyson:Dyson_series_F_W} of $F_W^{\alpha \alpha'}$. This approximation,
quadratic in $H_V$, is formally equivalent to a direct generalization of ND's solution
of the x-ray problem to several interacting quantum channels $\alpha$.
For conceptual details we refer to ND's article \cite{ND69}.

The central assumption in this approach consist in considering the localized state $\phi_d$
to be structureless. The scattering of conduction electrons described by $H_W$ therefore
depends on the occupancy of the $\phi_d$ state only, but on no other
internal structure of $\phi_d$. This allows us to consider the scattering potential
$W^{\alpha \alpha'}_{k k'}$ as a time-dependent external potential acting on the conduction
electrons during the times of occupancy of $\phi_d$.

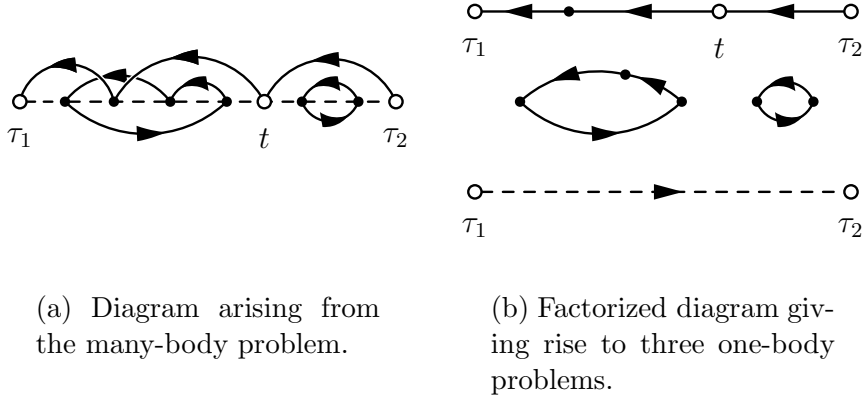
\begin{figure}
\begin{center}
	\subfigure[Diagram arising from the many-body problem.]
	{
		\begin{fmfgraph*}(50,40)
		    \fmfforce{(0,0.5h)}{i}		\fmfv{decor.shape=circle, decor.filled=empty, decor.size=5, l=$\tau_1$,l.a=-90,l.d=10}{i}
		    \fmfforce{(w,0.5h)}{o}		\fmfv{decor.shape=circle, decor.filled=empty, decor.size=5, l=$\tau_2$,l.a=-90,l.d=10}{o}
		    \fmfforce{(0.65w, 0.5h)}{t}	\fmfv{decor.shape=circle, decor.filled=empty, decor.size=5, l=$t$,     l.a=-90,l.d=10}{t}
		    \fmfforce{(0.12w,0.5h)}{v1}	\fmfv{decor.shape=circle, decor.filled=full, decor.size=3}{v1}
		    \fmfforce{(0.25w,0.5h)}{v2}	\fmfv{decor.shape=circle, decor.filled=full, decor.size=3}{v2}
		    \fmfforce{(0.40w,0.5h)}{v3}	\fmfv{decor.shape=circle, decor.filled=full, decor.size=3}{v3}
		    \fmfforce{(0.55w,0.5h)}{v4}	\fmfv{decor.shape=circle, decor.filled=full, decor.size=3}{v4}
		    \fmfforce{(0.75w,0.5h)}{v5}	\fmfv{decor.shape=circle, decor.filled=full, decor.size=3}{v5}
		    \fmfforce{(0.90w,0.5h)}{v6}	\fmfv{decor.shape=circle, decor.filled=full, decor.size=3}{v6}
		    \fmffreeze
		    \fmf {dashes,width=1}{i,v1}
		    \fmf {dashes,width=1}{v1,o}
		    \fmf {fermion, width=1, right=0.7}{v4,v3}
		    \fmf {fermion, width=1, right=0.5}{v3,v1}
		    \fmf {fermion, width=1, right=0.4}{v1,v4}
		    \fmf {fermion, width=1, right=0.7}{v6,v5}
		    \fmf {fermion, width=1, right=0.7}{v5,v6}
		    \fmf {fermion, width=1, right=.65}{o,t}
		    \fmf {fermion, width=1, right=.6,rubout}{t,v2}
		    \fmf {fermion, width=1, right=.8,rubout}{v2,i}
		\end{fmfgraph*}
		\label{fig:diagram_a}
	}
	\hspace{0.5cm}
	\subfigure[Factorized diagram giving rise to three one-body problems.]
	{
		\begin{fmfgraph*}(50,40)
		  \fmfforce{(0,0.8h)}{io}		\fmfv{decor.shape=circle, decor.filled=empty, decor.size=5, l=$\tau_1$,l.a=-90,l.d=10}{io}
		  \fmfforce{(w,0.8h)}{oo}		\fmfv{decor.shape=circle, decor.filled=empty, decor.size=5, l=$\tau_2$,l.a=-90,l.d=10}{oo}
		  \fmfforce{(0.65w, 0.8h)}{t}	\fmfv{decor.shape=circle, decor.filled=empty, decor.size=5, l=$t$,     l.a=-90,l.d=10}{t}
		  \fmfforce{(0.25w,0.8h)}{v2}	\fmfv{decor.shape=circle, decor.filled=full, decor.size=3}{v2}
		  \fmfforce{(0.12w,0.5h)}{v1}	\fmfv{decor.shape=circle, decor.filled=full, decor.size=3}{v1}
		  \fmfforce{(0.40w,0.59h)}{v3}	\fmfv{decor.shape=circle, decor.filled=full, decor.size=3}{v3}
		  \fmfforce{(0.55w,0.5h)}{v4}	\fmfv{decor.shape=circle, decor.filled=full, decor.size=3}{v4}
		  \fmfforce{(0.75w,0.5h)}{v5}	\fmfv{decor.shape=circle, decor.filled=full, decor.size=3}{v5}
		  \fmfforce{(0.90w,0.5h)}{v6}	\fmfv{decor.shape=circle, decor.filled=full, decor.size=3}{v6}
		  \fmfforce{(0,0.2h)}{iu}		\fmfv{decor.shape=circle, decor.filled=empty, decor.size=5, l=$\tau_1$,l.a=-90,l.d=10}{iu}
		  \fmfforce{(w,0.2h)}{ou}		\fmfv{decor.shape=circle, decor.filled=empty, decor.size=5, l=$\tau_2$,l.a=-90,l.d=10}{ou}
		  \fmffreeze
		  \fmf {dashes_arrow,width=1}{iu,ou}
		  \fmf {fermion, width=1, right=0.15}{v4,v3}
		  \fmf {fermion, width=1, right=0.25}{v3,v1}
		  \fmf {fermion, width=1, right=0.4}{v1,v4}
		  \fmf {fermion, width=1, right=0.7}{v6,v5}
		  \fmf {fermion, width=1, right=0.7}{v5,v6}
		  \fmf {fermion, width=1}{oo,t}
		  \fmf {fermion, width=1}{t,v2}
		  \fmf {fermion, width=1}{v2,io}
		\end{fmfgraph*}
		\label{fig:diagram_b}
	}
	\caption{Example of the factorization of the many-body problem. The plain lines represent free
	conduction electron propagators (evolving under $H_0$) and the dashed line the propagator
	of the particle in $\phi_d$. The full dots are interaction vertices due to $W^{\alpha \alpha'}$
	and the open dots at $\tau_1$ and $\tau_2$ due to $V^\alpha$.}
	\label{fig:lowest:example}
\end{center}
\end{figure}

At the order $n=1$, for fixed times $\tau_1$ and $\tau_2$ with, for instance, $\tau_1 < \tau_2$,
the hybridization term $H_V$ at the time $\tau_1$ describes the tunneling of an electron into $\phi_d$,
whereas the hybridization term at $\tau_2$ transfers this particle back to one of the electrodes.
The potential $W^{\alpha \alpha'}_{k k'}$ applies therefore during the interval $(\tau_1,\tau_2)$
and is zero for all other times. 
The time $t$ must lie between $\tau_1$ and $\tau_2$ because of the operators $d^\dagger(t) d(t)$
in the definition of $F^{\alpha \alpha'}_W(t)$.
The only contribution of the localized state to the amplitude is its free propagation factor
\[
	\exp\bigl( -i E_d (\tau_2-\tau_1) \bigr).
\]
If we expand the amplitude 
$\bra{i} T \{ \widehat{H}_V(\tau_1) \widehat{H}_V(\tau_2) 
 \widehat{C}^\dagger_\alpha(t) \widehat{C}_\alpha'(t) \} \ket{i}$
in the external potential, the Feynman diagrams arising from the different contractions factorize into 
two groups: the diagrams that are connected to at least one of the points $\tau_1,\tau_2$ and $t$,
and the closed loops where no vertex coincides with one of these points (see Fig.~\ref{fig:lowest:example}
for an example).
If we denote by $L^{\alpha \alpha'}(\tau_1,\tau_2,t)$ the former group of diagrams, and by 
$\e^{C(\tau_1,\tau_2)}$ the latter, where by the linked cluster theorem the complete
closed loop sum can be expressed by the exponential of the sum $C(\tau_1,\tau_2)$ of all
single closed loops, the amplitude at $n=1$ in \eqref{eq:dyson:Dyson_series_F_W} becomes
\begin{equation} \label{eq:lowest:decomp}
	\bra{i} T \bigl\{ \widehat{H}_V(\tau_1) \widehat{H}_V(\tau_2) 
 	\widehat{C}^\dagger_\alpha(t) \widehat{C}_{\alpha'}(t) \bigr\} \ket{i} = 
	L^{\alpha \alpha'}(\tau_1,\tau_2,t) \, \e^{C(\tau_1,\tau_2)} \, \e^{-i E_d (\tau_2-\tau_1)}
\end{equation}

%----------------------------------------------------------------------------------------------------------

As we will see below, both quantities, $L^{\alpha \alpha'}(\tau_1,\tau_2,t)$ and $C(\tau_1,\tau_2)$, 
can be expressed by one-body 
propagators $G^{\alpha \alpha'}(\tau,\tau')$ describing a particle which is created at the time $\tau'$
in the electrode $\alpha'$ and annihilated at the time $\tau$ in the electrode $\alpha$, and which propagates under
the external potential $W$ during the time interval $(\tau_1,\tau_2)$,
\begin{equation} \label{eq:lowest:def_Greens_fct}
	G^{\alpha \alpha'}(\tau,\tau') = 
		\bra{i} T \bigl\{ \widehat{C}_\alpha(\tau) \widehat{C}^\dagger_{\alpha'}(\tau')
		\bigr\} \ket{i}.
\end{equation}
This propagator is determined by the Dyson equation
\begin{equation} \label{eq:lowest:Dyson_eq}
	G^{\alpha \alpha'}(\tau,\tau') = G^\alpha_0(\tau,\tau') \delta_{\alpha \alpha'}
	- i \sum_{\alpha''} \int_{\tau_1}^{\tau_2}\dtau'' \, G^\alpha_0(\tau,\tau'') \, W^{\alpha \alpha''}
	\, G^{\alpha'' \alpha'}(\tau'',\tau')
\end{equation}
where $G^\alpha_0(\tau,\tau')$ is the free propagator.
The factor $(-i)$ in front of the integral in \eqref{eq:lowest:Dyson_eq} arises from the unusual definition 
\eqref{eq:lowest:def_Greens_fct} of the Green's functions.
The Dyson equation will be calculated using the asymptotic expressions for large $|\tau-\tau'|$ developed
by ND. Since for the free propagators, the electrodes $\alpha$ are assumed to be independent,
ND's development can be carried out for each $G^\alpha_0$ separately, and is given by
(see \cite{ND69}, Eq.~(38))
\begin{equation} \label{eq:lowest:free_prop}
	G^\alpha_0(\tau-\tau') = (-i \nu_\alpha) \left[ 
		\ppf{1}{\tau-\tau'} + 
		\pi \tan\vartheta_\alpha \, \delta(\tau-\tau') 
	\right] \, \e^{-i \mu_\alpha (\tau-\tau')}
\end{equation}
where the $\mathcal{P}$ indicates a principal value distribution. The angle $\tan\vartheta_\alpha$ 
contains the short-time corrections to the asymptotic form in $1/(\tau-\tau')$ and depends on the
form of the conduction band. For a completely symmetric band around the Fermi surface it vanishes.
If we introduce the propagators \eqref{eq:lowest:free_prop} into the Dyson equation \eqref{eq:lowest:Dyson_eq}
and integrate over the
delta functions, we find the following system of singular integral equations
\begin{equation} \label{eq:lowest:Singular_Integral}
	\widetilde{A}(\tau) G(\tau,\tau') - \frac{1}{\pi i} \int_{t_0}^{t_1} \dtau'' \, 
	\ppf{1}{\tau'' - \tau}
	\widetilde{B}(\tau'') \, G(\tau'',\tau') = \widetilde{f}(\tau,\tau').
\end{equation}
where the coefficients $\widetilde{A}$ and $\widetilde{B}$ are $2 \times 2$ matrix functions of the form
\begin{align}
	\widetilde{A}(\tau) &\equiv \e^{i \mu \tau} A 
	                     = \e^{i \mu \tau} (\openone + \pi \tan\vartheta g),  \label{eq:lowest:A} \\
	\widetilde{B}(\tau) &\equiv \e^{i \mu \tau} B              
	                     = \e^{i \mu \tau} (i \pi g),                         \label{eq:lowest:B} 
\end{align}
for the $2 \times 2$ matrices $\mu = \{\mu_\alpha \delta_{\alpha \alpha'}\}$,
$\tan\vartheta = \{\tan\vartheta_\alpha \delta_{\alpha \alpha'}\}$, and
$g \equiv \nu W = \{ \nu_\alpha W^{\alpha \alpha'} \}$, and where $\openone$ is the $2 \times 2$
unit matrix.
The inhomogeneous term $\widetilde{f}(\tau,\tau')$ is given by   
\begin{equation} \label{eq:lowest:f} 
	\widetilde{f}(\tau,\tau') = \e^{i \mu \tau} G_0(\tau-\tau'). 
\end{equation}
The system of singular integral equations \eqref{eq:lowest:Singular_Integral} turns out to be exactly 
solvable. We provide the solution in the Appendix~\ref{sec:int}.
Let us define the constant matrices
\begin{align}
	S &= A + B = \openone + \pi ( \tan\vartheta + i \openone) g, \label{eq:lowest:S}  \\
	D &= A - B = \openone + \pi ( \tan\vartheta - i \openone) g. \label{eq:lowest:D} 
\end{align}	
As shown in Appendix~\ref{sec:unitary}, the matrix
\begin{equation}  \label{eq:lowest:S_matrix}
 	\mathfrak{S} = S D^{-1} = \bigl( \openone - \pi ( \tan\vartheta + i \openone) g \bigr)
 	                         \bigl[ \openone - \pi ( \tan\vartheta - i \openone) g \bigr]^{-1}
\end{equation}
is the inverse of the (unitary) scattering matrix due to $W^{\alpha \alpha'}$ evaluated on the chemical 
potentials $\mu_\alpha$.
Thus, if we write
\begin{equation}
	\mathfrak{S} = \e^{2 i \Delta},
\end{equation}
the solution of the Dyson equation \eqref{eq:lowest:Dyson_eq}, for $\tau, \tau' \in (\tau_1,\tau_2)$ 
is given by Eq.~\eqref{eq:int_eq:solution_Singular_Integral}, and reads
\begin{multline} \label{eq:lowest:solution_Singular_Integral}
	G(\tau,\tau') = (-i) \, S^{-1}(\tau) \e^{-i \Delta(\tau)} 
		\left| \frac{\tau-\tau_2}{\tau-\tau_1} \right|^{\frac{\Delta}{\pi}} \e^{-i \mu (\tau-\tau')} 
		\left| \frac{\tau'-\tau_1}{\tau'-\tau_2} \right|^{\frac{\Delta}{\pi}} \e^{+i \Delta(\tau)} \\
		\left[ \ppf{1}{\tau-\tau'} + \pi \Theta(\tau') \, \delta(\tau-\tau') \right] \,
		M(\tau') \nu,
\end{multline}
for the matrices
\begin{align} 
	\Theta(\tau') &= \begin{cases}
		i ( S D - A ) B^{-1}	&	\text{if $\tau' \in (\tau_1,\tau_2)$,} \\
		\tan\vartheta			&	\text{otherwise,}		\label{eq:lowest:Theta} 
	\end{cases} \\
	M(\tau') &= \begin{cases}
		B D^{-1} B^{-1}			&	\text{if $\tau' \in (\tau_1,\tau_2)$,} \\
		\openone				&	\text{otherwise,}		\label{eq:lowest:M} 
	\end{cases} \\
	S(\tau) &= \begin{cases}
		S						&	\text{if $\tau  \in (\tau_1,\tau_2)$,} \\
		\openone				&	\text{otherwise,}		\label{eq:lowest:S(tau)} 
	\end{cases} \\
	\Delta(\tau) &= \begin{cases}
		\Delta					&	\text{if $\tau  \in (\tau_1,\tau_2)$,} \\
		0 						&	\text{otherwise.}		\label{eq:lowest:Delta(tau)} 
	\end{cases}
\end{align}
In the limit where the non-diagonal entries of $W^{\alpha \alpha'}$ vanish and the electrodes are
decoupled, we recover for $G^{\alpha \alpha'}(\tau,\tau')$ ND's result (\cite{ND69}, Eq. \mbox{(51)}).

%----------------------------------------------------------------------------------------------------------

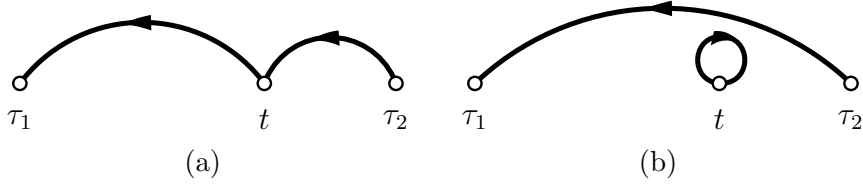
\begin{figure}
\begin{center}
	\subfigure[]
	{
		\begin{fmfgraph*}(50,15)
			\fmfforce{(0,0.2h)}{i}      \fmfv{decor.shape=circle, decor.filled=empty, decor.size=5, l=$\tau_1$,l.a=-90,l.d=10}{i}
			\fmfforce{(w,0.2h)}{o}      \fmfv{decor.shape=circle, decor.filled=empty, decor.size=5, l=$\tau_2$,l.a=-90,l.d=10}{o}
			\fmfforce{(0.65w, 0.2h)}{t} \fmfv{decor.shape=circle, decor.filled=empty, decor.size=5, l=$t$,     l.a=-90,l.d=10}{t}
			\fmffreeze
			\fmf {fermion,width=2, right=0.7}{o,t}
			\fmf {fermion,width=2, right=0.5}{t,i}
		\end{fmfgraph*}
		\label{fig:diagram_La}
	}
	\hspace{0.5cm}
	\subfigure[]
	{
		\begin{fmfgraph*}(50,15)
			\fmfforce{(0,0.2h)}{i}       \fmfv{decor.shape=circle, decor.filled=empty, decor.size=5, l=$\tau_1$,l.a=-90,l.d=10}{i}
			\fmfforce{(w,0.2h)}{o}       \fmfv{decor.shape=circle, decor.filled=empty, decor.size=5, l=$\tau_2$,l.a=-90,l.d=10}{o}
			\fmfforce{(0.65w, 0.2h)}{v1} \fmfv{decor.shape=circle, decor.filled=empty, decor.size=5, l=$t$,     l.a=-90,l.d=10}{v1}
			\fmfforce{(0.66w, 0.2h)}{v2}	 
			\fmffreeze
			\fmf {fermion,width=2, right=0.4}{o,i}
			\fmf {fermion,width=2,left=25}{v1,v2}
		\end{fmfgraph*}
		\label{fig:diagram_Lb}
	}
	\caption{The two possible contractions contributing to $L(\tau_1,\tau_2,t)$. The bold
	lines represent the propagator $G$, given in \eqref{eq:lowest:solution_Singular_Integral}.}
	\label{fig:diagram_L}
\end{center}
\end{figure}

Let us now calculate the factor $L^{\alpha \alpha'}(\tau_1,\tau_2,t)$. Since we know the propagator 
$G(\tau,\tau')$ in the presence of $W$, we will use Wick's theorem to contract the four conduction
electron operators at the points $\tau_1, \tau_2$ and $t$. Every contraction is associated
to a propagator $G(\tau,\tau')$. 
As we see in Fig.~\ref{fig:diagram_L}, there are only two possible diagrams.
The first diagram, Fig.~\ref{fig:diagram_La}, contributes
\[
	- (-1)^2 \sum_{\alpha_1 \alpha_2} G^{\alpha_1 \alpha}(\tau_1,t) \, G^{\alpha' \alpha_2}(t,\tau_2)
\]
where a contraction of the operators $\widehat{C}^\dagger(\tau') \widehat{C}(\tau)$
equals $-G(\tau,\tau')$ and where the third minus sign arises from the commutation of
$\widehat{C}^\dagger_\alpha(t)$ and $\widehat{C}_{\alpha'}(t)$ to separate the contractions.
The second diagram, Fig.~\ref{fig:diagram_Lb} gives, accordingly,
\[
	(-1)^2 \sum_{\alpha_1 \alpha_2} G^{\alpha_1 \alpha_2}(\tau_1,\tau_2) \, G^{\alpha' \alpha}(t,t).
\]
For convenience, we introduce the auxiliary times $\tau_3 = \tau_4 = t$, and the indices
$\alpha_3 = \alpha'$ and $\alpha_4 = \alpha$. With these notations we find that
\begin{equation} \label{eq:lowest:L}
	L^{\alpha \alpha'}(\tau_1, \tau_2, t) = \sum_{\alpha_1 \alpha_2} V^{\alpha_1} (V^{\alpha_2})^* \ 
		\det G^{\alpha_{2j-1} \alpha_{2k}}(\tau_{2j-1},\tau_{2k})
\end{equation}
where the determinant is taken over the indices $j,k = 1,2$.

If we introduce the explicit form  \eqref{eq:lowest:solution_Singular_Integral} of the propagator
into the latter equation, we are lead to divergences in both propagators $G(\tau_2,\tau_1)$ and
$G(t,t)$. 
According to ND, these divergences are due to the asymptotic approximation 
of the propagators and must be cured by a (pure imaginary) cutoff $i \xi_0$ 
of the order of the bandwidth, such that
\begin{multline} \label{eq:lowest:G_cutoffed}
	G(\tau_1,\tau_2) = (-i) \, \frac{1}{\tau_1-\tau_2} \, S^{-1} \, \e^{-i \Delta} \, 
	\bigl( i \xi_0 |\tau_2 - \tau_1| \bigr)^{-\frac{\Delta}{\pi}} \,
	\e^{- i \mu (\tau_1 - \tau_2) }    \\
	\e^{+i \Delta} \, 
	\bigl( i \xi_0 |\tau_2 - \tau_1| \bigr)^{-\frac{\Delta}{\pi}} \,
	B D^{-1} B^{-1} \nu, 
\end{multline}
and, if we interpret $G(t,t)$ as $G(t-i\xi_0^{-1},t)$,
\begin{equation} \label{eq:lowest:G(t,t)}
	G(t,t) = (-i) \ S^{-1} B D^{-1} B^{-1} \,  \nu \ (i\xi_0).
\end{equation}
Notice that in this form, $G(t,t)$ does not depend explicitly on the boundaries $\tau_1$ and $\tau_2$.
This might be an acceptable approximation for the Green's function in the determinant of $L^{\alpha \alpha'}$,
but it is not sufficient if we consider the closed loop sum.

%----------------------------------------------------------------------------------------------------------

The sum over all simple closed loops can be calculated by use of the equation (\cite{ND69}, Eq.~(22))
\begin{equation}
	\lambda \frac{\partial C(\tau_1,\tau_2)}{\partial \lambda} \biggr|_{\lambda = 1}
	= i \sum_{\alpha \alpha'} \int_{\tau_1}^{\tau_2} \dtau \, W^{\alpha \alpha'} \, G^{\alpha' \alpha}(\tau,\tau)
	= i \int_{\tau_1}^{\tau_2} \dtau \ \Tr\left[ W \, G(\tau,\tau) \right]
\end{equation}
where in $C(\tau_1,\tau_2)$ all potentials $W$ have been multiplied by the parameter $\lambda$.

Again $G(\tau,\tau)$ is singular, but if we develop $G(\tau,\tau')$ in powers of $\Delta\tau = \tau-\tau'$,
we can separate out the singular term which diverges like $1/\Delta\tau$. Namely, since
\begin{multline} \label{eq:lowest:development_X}
	X(\tau) \, \e^{-i \mu \Delta\tau} \, X^{-1}(\tau') 
	= \left[ \frac{\tau-\tau_2}{\tau-\tau_1}\right]^{\frac{\Delta}{\pi}} 
		\e^{-i \mu \Delta\tau} \\
		\left\{ \left[\frac{\tau - \tau_1}{\tau-\tau_2}\right]^{\frac{\Delta}{\pi}} +
			\frac{\Delta}{\pi} \left[\frac{\tau - \tau_1}{\tau-\tau_2}\right]^{\frac{\Delta}{\pi}-1} 
			\left( \frac{\tau - \tau_1}{(\tau-\tau_2)^2} - \frac{1}{\tau-\tau_2} \right) \Delta\tau 
			+ \mathcal{O}(\Delta\tau)^2 
		\right\}
\end{multline}	
we have in the limit $\Delta\tau \to 0$
\begin{equation} \label{eq:lowest:G(tau,tau)}
	G(\tau,\tau') \xrightarrow[\Delta\tau \to 0]{}	
	\text{(Sing.)} - (-i) \left[\frac{1}{\tau_2-\tau} + \frac{1}{\tau-\tau_1}\right] \ 
	S^{-1}(\tau) \frac{\Delta}{\pi} M(\tau) \nu
\end{equation}
where the singular part (Sing.) is due only to the approximation on the propagators.
The important point is that it does not depend on $\tau_1$ and $\tau_2$.
As shown in ND, its contribution to the exponential is proportional to the first correction
in energy (for a static potential $W$ extending in time from $-\infty$ to $+\infty$)
which will be denoted by $\Delta E_1$. 

The integration of $\tau$ over the interval $(\tau_1,\tau_2)$ leads to a logarithmic divergence at the boundaries
and requires again the introduction of the cutoff $\xi_0$, so that 
(considering only the regular parts of $G(\tau,\tau)$)
\begin{equation}
\begin{split}
	\lambda \frac{\partial C}{\partial \lambda}\biggl|_{\lambda=1} 
	&=  \frac{2}{\pi} \, \log\big[i |\tau_2-\tau_1| \xi_0\big] \
		\Tr\left[ W S^{-1} \Delta B D^{-1} B^{-1} \nu \right] \\
	&=  \frac{2i}{\pi^2} \, \log\big[i |\tau_2-\tau_1| \xi_0\big] \
		\Tr\left[ \Delta B D^{-1} S^{-1}\right] \\
	&=  - \frac{1}{\pi^2} \, \log\big[i |\tau_2-\tau_1| \xi_0\big] \
		\lambda \, \frac{\mathrm{d}}{\mathrm{d}\lambda} \Tr\left[ \Delta^2\right].
\end{split}
\end{equation}		
The last equality can be shown as follows:
If $M(\lambda)$ is an invertible matrix depending on the parameter $\lambda$, we have
\begin{equation} \label{eq:lowest:derivative_M^-1}
	\frac{\mathrm{d}}{\mathrm{d}\lambda}M^{-1}(\lambda) = 
	- M^{-1}(\lambda) \left(\frac{\mathrm{d}}{\mathrm{d}\lambda}M(\lambda)\right) M^{-1}(\lambda),
\end{equation}
as can easily be verified by deriving the matrix product $M(\lambda) M^{-1}(\lambda) = \openone$ with
respect to $\lambda$.
Therefore, if we replace the potential $W$ by $\lambda W$ in the definition of the matrix
$\mathfrak{S}$ in Eq.~\eqref{eq:lowest:S_matrix}, and derive $\mathfrak{S}^{-1}$ 
with respect to $\lambda$ by using \eqref{eq:lowest:derivative_M^-1}, we find after some
algebra
\[
	\lambda \mathfrak{S} \frac{\mathrm{d} \mathfrak{S}^{-1} }{\mathrm{d}\lambda}
	= - 2 B D^{-1} S^{-1}.
\]
Furthermore, since $\Delta$ is defined by $\mathfrak{S} = \e^{2 i \Delta}$, we have
\begin{multline*}
	(-2) \Tr\left[ \Delta B D^{-1} S^{-1} \right]
	= \lambda \Tr\left[\Delta \mathfrak{S} \frac{\mathrm{d}}{\mathrm{d} \lambda} \mathfrak{S}^{-1}\right]\\ 
	= \lambda \Tr\left[\Delta \e^{2 i \Delta} \frac{\mathrm{d}}{\mathrm{d} \lambda} \e^{-2i\Delta}\right] 
	= - i \lambda \frac{\mathrm{d}}{\mathrm{d} \lambda} \Tr\left[ \Delta^2 \right]
\end{multline*}
where we have used the invariance of the trace under cyclic permutations.
With help of the latter expression we can integrate the simple closed loop sum. This yields,
reintroducing the contribution in $\Delta E_1$,
\begin{equation}
	C = \int_0^1 \mathrm{d}\lambda \, \frac{\mathrm{d} C}{\mathrm{d} \lambda}
	  = - \frac{1}{\pi^2} \, \log\left[ i |\tau_2-\tau_1| \xi_0 \right] \, \Tr\left[\Delta^2\right]
	  	- i \Delta E_1 \, (\tau_2-\tau_1)
\end{equation}
and the closed loops contribute
\begin{equation} \label{eq:lowest:closed_Loops}
	\e^{C(\tau_1,\tau_2)} = \bigl[i|\tau_2-\tau_1|\xi_0\bigr]^{- \frac{\Tr \Delta^2}{\pi^2}} \ 
					\e^{-i \Delta E_1 (\tau_2-\tau_1)}.
\end{equation}

%----------------------------------------------------------------------------------------------------------

Putting together Eqs.~\eqref{eq:lowest:L} and \eqref{eq:lowest:closed_Loops}, the contribution
to $F_W^{\alpha \alpha'}(t)$ in the series \eqref{eq:dyson:Dyson_series_F_W} takes the form
\begin{multline} \label{eq:lowest:amplitude}
	\bigr[F^{\alpha \alpha'}_W(t)\bigr]_{(2)}(-i)^2  =
	\underset{t_0 < \tau_1 < t < \tau_2 < t_1}{\int} \dtau_1 \dtau_2 \
		L^{\alpha \alpha'}(\tau_1,\tau_2,t) \, \e^{C(\tau_1,\tau_2)} \, \e^{-i E_d (\tau_2-\tau_1)} 
	=  \\
	(-i)^2 \underset{t_0 < \tau_1 < t < \tau_2 < t_1}{\int} \dtau_1 \dtau_2 \, 
		\sum_{\alpha_1 \alpha_2} (V^{\alpha_1})^* V^{\alpha_2} \ 
		\det G^{\alpha_{2j-1} \alpha_{2k}}(\tau_{2j-1},\tau_{2k}) \, \\
		\bigl[i \xi_0 |\tau_2-\tau_1|\bigr]^{-\frac{\Tr \Delta^2}{\pi^2}} \, \e^{-i (E_d+\Delta E_1) (\tau_2-\tau_1)}
\end{multline}
where $\tau_3 = \tau_4 = t$, $\alpha_3 = \alpha'$, $\alpha_4 = \alpha$, and 
$j,k=1,2$. The factor $1/2!$ occurring in \eqref{eq:dyson:Dyson_series_F_W}
has been eliminated by the restriction of the domain of integration.

%-----------------------------------------------------------------------------------------------------

\section{Extension to higher orders}
\label{sec:higher}

An extension to higher orders in perturbation of  ND's ideas was carried out already a long time
ago by Yuval and Anderson \cite{YA70} who derived in this way an expression for the partition function 
of the single spin $1/2$ impurity Kondo problem. In this section we show how their method
can be generalized to the present problem. Consider an amplitude at order $n \ge 2$ in the Dyson
series \eqref{eq:dyson:Dyson_series_F_W} for $F^{\alpha \alpha'}_W$. If the intermediate times $\tau_j$
are fixed and chosen such that $\tau_1 < \tau_2 < \dots < \tau_{2n}$, an electron tunnels into
the localized state at odd times $\tau_{2j-1}$, and leaves it at even times $\tau_{2j}$ ($j=1,\dots,n$).
As before, these times have to be chosen such that the state $\phi_d$ is occupied at the time $t$.
The many-body problem can be factorized into one-body problems as in the previous section, and the 
amplitude in the expansion \eqref{eq:dyson:Dyson_series_F_W} can be written in a form similar to 
Eq.~\eqref{eq:lowest:decomp},
\begin{multline} \label{eq:higher:decomp}
	\bra{i} T \bigl\{ \widehat{H}_V(\tau_1) \dots \widehat{H}_V(\tau_{2n}) \, 
		\widehat{C}^\dagger_{\alpha}(t) \, \widehat{C}_{\alpha'}(t) 
	\bigr\} \ket{i} = \\
	L^{\alpha \alpha'}(\tau_1,\dots, \tau_{2n},t) \, \e^{C(\tau_1,\dots,\tau_{2n})} \, 
	\e^{-i E_d \sum_{j=1}^n (\tau_{2j}-\tau_{2j-1})},
\end{multline}
where $L^{\alpha \alpha'}(\tau_1,\dots,\tau_{2n},t)$ contains all diagrams that are connected to
at least one of the times $\tau_1,\dots,\tau_{2n}$ or $t$, and $\e^{C(\tau_1,\dots,\tau_{2n})}$
is the closed loop sum. Both quantities are evaluated in the external potential 
$W(\tau)$ which equals $W$ for $\tau$ in the intervals 
$(\tau_{2j-1},\tau_{2j}), j = 1,\dots,n$, and is zero for all other times.
The one-particle propagator in this potential obeys the Dyson equation
\begin{equation} \label{eq:higher:Dyson_eq}
	G(\tau,\tau') = G_0(\tau-\tau') - i \int_{t_0}^{t_1} \dtau'' G_0(\tau-\tau'') W(\tau'') G(\tau'',\tau').
\end{equation}
The solution of the latter equation is given in Eq.~\eqref{eq:int_eq:solution_Singular_Integral_b} and
reads
\begin{multline} \label{eq:higher:solution_Singular_Integral}
	G(\tau,\tau') = (-i) \, S^{-1}(\tau) \e^{-i \Delta(\tau)} 
		\left| \prod_{j=1}^n \frac{\tau-\tau_{2j}}{\tau-\tau_{2j-1}} \right|^{\frac{\Delta}{\pi}} 
			\e^{-i \mu (\tau-\tau')} 
		\left| \prod_{j=1}^n \frac{\tau'-\tau_{2j-1}}{\tau'-\tau_{2j}} \right|^{\frac{\Delta}{\pi}}
			\e^{+i \Delta(\tau)} \\
		\left[ \ppf{1}{\tau-\tau'} + \pi \Theta(\tau') \, \delta(\tau-\tau') \right] \,
		M(\tau') \nu.
\end{multline}
for the same definitions of the matrices as in Eqs.~\eqref{eq:lowest:Theta}--\eqref{eq:lowest:S(tau)}
in which the conditions $\tau \in (\tau_1,\tau_2)$ are replaced by 
$\tau \in (\tau_{2j-1},\tau_{2j}), j = 1,\dots, n$.

In order to determine the factor $L^{\alpha \alpha'}(\tau_1,\dots,\tau_{2n},t)$ we take advantage from the 
fact that the $\tau_j$ are time ordered, $\tau_1 < \dots < \tau_{2n}$. A full contraction of the
operators at the times $\tau_1,\dots,\tau_{2n},\tau_{2n+1}=t,\tau_{2n+2} = t$ 
can be regarded as a permutation $\pi$ of $n+1$ elements, where the creator
$\widehat{C}^\dagger(\tau_{2j})$ is contracted with the annihilator $\widehat{C}(\tau_{2\pi(j)-1})$.
By Wick's theorem, the whole amplitude must be multiplied by the signature of the permutation that
brings all contracted operators together. Since the operators at the times $\tau_1, \dots, \tau_{2n}$ 
are explicitly time ordered and since the operators
at $t$ occur as a pair of creation and annihilation operators (for which their relative position
with respect to the $\tau_j$ is of no importance) this separation is 
performed by $\pi^{-1}$. 
Every contraction is expressed by the factor 
$-G^{\alpha_{2\pi(j)-1},\alpha_{2j}}(\tau_{2\pi(j)-1},\tau_{2j})$ as was seen in the previous section.
Therefore, the sum over all contractions equals the sum over all permutation $\pi$ and is given by
($\tau_{2n+1}=\tau_{2n+2}=t, \alpha_{2n+1}=\alpha', \alpha_{2n+2}=\alpha$)
\begin{multline}
	L^{\alpha \alpha'}(\tau_1,\dots,\tau_{2n},t) 
	= \\
	(-1)^n \sum_{\alpha_1 \dots \alpha_{2n}} 
		(V^{\alpha_1})^* V^{\alpha_2} \dots V^{\alpha_{2n}}
		\sum_\pi \text{sign}(\pi^{-1}) 
			\prod_{j=1}^n G^{\alpha_{2\pi(j)-1},\alpha_{2j}}(\tau_{2\pi(j)-1},\tau_{2j}) 
	= \\
	(-1)^n \sum_{\alpha_1 \dots \alpha_{2n}} 
		(V^{\alpha_1})^* V^{\alpha_2} \dots V^{\alpha_{2n}}
		\det{}_{n+1} G^{\alpha_{2j-1},\alpha_{2k}}(\tau_{2j-1},\tau_{2k})
\end{multline}
where we used Leibniz' definition of the determinant. The subscript $n+1$ in the determinant
indicates that it must be taken over the indices 
$j,k = 1, \dots, n+1$. Again the Green's functions $G(\tau_{2j-1},\tau_{2k})$ and $G(t,t)$ are singular
and must be cured by the cutoff $\xi_0$ as in Eqs.~\eqref{eq:lowest:G_cutoffed} and \eqref{eq:lowest:G(t,t)}.
The closed loops are determined by the equation
\begin{equation}
	\lambda \frac{\partial C(\tau_1,\dots,\tau_{2n})}{\partial \lambda} \biggl|_{\lambda=1}
	= i \int_{t_0}^{t_1} \dtau \ \Tr\left[ W(\tau) \, G(\tau,\tau) \right].
\end{equation}
A similar kind of investigation of the singular and nonsingular behavior of $G(\tau,\tau)$
as in Eq.~\eqref{eq:lowest:G(tau,tau)} leads to the simple closed loop sum
\begin{equation} \label{eq:higher:closed_loops}
	C(\tau_1,\dots,\tau_{2n}) = -i \Delta E_1 \sum_{j=1}^n (\tau_{2j}-\tau_{2j-1}) + 
		\frac{\Tr \Delta^2}{\pi^2} \sum_{\substack{i,j=1\\i<j}}^{2n} 
		(-1)^{i+j} \, \ln\bigl(i \xi_0 |\tau_j - \tau_i|\bigr)
\end{equation}	
In order to obtain the full expression at order $2n$ for $F^{\alpha \alpha'}_W(t)$, we must
integrate over the times $\tau_1, \dots, \tau_{2n}$ with the constraint of $t$ lying on a time
interval in which the localized state is occupied. This can be taken into account 
if we multiply the integrand by the expression 
$\sum_{m=1}^{n+1} \delta_{\alpha_{2m},\alpha'} \, \delta_{\alpha_{2m+1},\alpha} \, \delta(t-\tau_{2m}) \, 
\delta(t-\tau_{2m+1})$ 
and integrate over the $2n+2$ time variables $\tau_1 < \dots < \tau_{2n+2}$ and sum over the indices
$\alpha_1, \dots, \alpha_{2n+2}$. The time order of the integration variables drops the factor
$1/(2n)!$ in the Dyson series, and the full amplitude reads
\begin{multline} \label{eq:higher:amplitude}
	\bigl[F^{\alpha \alpha'}_W(t)\bigr]_{(2n)} = 
	(-i)^n  \underset{t_0 < \tau_1 < \dots < \tau_{2n+2} < t_1}{\int} \dtau_1 \dots \dtau_{2n+2} \,
	\sum_{\alpha_1 \dots \alpha_{2n}} 
	(V^{\alpha_1})^* V^{\alpha_2} \dots V^{\alpha_{2n+2}} \\
	\frac{1}{(V^\alpha)^* V^{\alpha'}} \Bigl[ \sum_{m=1}^{n+1}
	\delta_{\alpha_{2m},\alpha'} \, \delta_{\alpha_{2m+1},\alpha} \, \delta(t-\tau_{2m}) \, 
	\delta(t-\tau_{2m+1}) \Bigr] \
	\det{}_{n+1} G^{\alpha_{2j-1} \alpha_{2k}}(\tau_{2j-1},\tau_{2k}) \\
	\e^{-i (E_d+\Delta E_1) \sum_{j=1}^n (\tau_{2j}-\tau_{2j-1})} \
	\exp\Bigl[	
		\tfrac{\Tr \Delta^2}{\pi^2} \sum_{\substack{i,j=1\\i<j}}^{2n} (-1)^{i+j} \, \ln\bigl(i \xi_0 |\tau_j - \tau_i|\bigr)
	\Bigr].
\end{multline}

%-----------------------------------------------------------------------------------------------------
\section{Discussion and limiting cases}
\label{sec:disc}

If we introduce the expressions \eqref{eq:lowest:amplitude} and \eqref{eq:higher:amplitude}
into the Dyson series \eqref{eq:dyson:Dyson_series_F_W} for $F^{\alpha \alpha'}_W$, we obtain
\begin{multline} \label{eq:disc:F_W}
	(V^\alpha)^* V^{\alpha'} \, F^{\alpha \alpha'}_W(t) = 
	\sum_{n=2}^{\infty} \, (-i)^{2n} \underset{t_0 < \tau_1 < \dots < \tau_{2n} < t_1}{\int}
	\dtau_1 \dots \dtau_{2n} \, \sum_{\alpha_1 \dots \alpha_{2n}} \\
	\sum_{m=1}^n \Bigl[ \delta_{\alpha_{2m-1},\alpha} \, \delta_{\alpha_{2m},\alpha'} \, 
	\delta(\tau_{2m-1}-t) \, \delta(t-\tau_{2m}) \Bigr] \ 
	(V^{\alpha_1})^* V^{\alpha_2} \dots V^{\alpha_{2n}} \\
	\det{}_{n} G^{\alpha_{2j-1} \alpha_{2k}}(\tau_{2j-1},\tau_{2k}) \,
	\exp\Bigl[\tfrac{\Tr \Delta^2}{\pi^2}\sum_{\substack{i,j=1\\i<j}}^{2n} (-1)^{i+j} \, 
		\ln\bigl(i \xi_0 (\tau_j-\tau_i) \bigr)\Bigr] \\
	\e^{-i (E_d+\Delta E_1) \sum_{j=1}^n (\tau_{2j}-\tau_{2j-1})}.
\end{multline}
A similar derivation holds for $F^\alpha_V(t)$. The main difference lies in the fact
that one single creation operator is fixed at the time $t$. Its relative position with respect
to the times of integration is important to determine the signature of the permutation separating
the contractions. We express this property by multiplying the integrand by 
$\sum_{m=1}^n \delta_{\alpha, \alpha_{2m}} \delta(t-\tau_{2m})$. The integration is performed
over $\tau_1 < \dots < \tau_{2n}$, and the sum over $\alpha_1, \dots, \alpha_{2n}$. This yields
\begin{multline} \label{eq:disc:F_V}
	V^\alpha F_V^\alpha(t) = \sum_{n=1}^\infty (-i)^{2n-1} 
	\underset{t_0 < \tau_1 < \dots < \tau_{2n} < t_1}{\int} \dtau_1 \dots \dtau_{2n} 
	\Bigl[ \sum_{m=1}^n
	\delta_{\alpha, \alpha_{2m}} \delta(t-\tau_{2m}) \Bigr]\\
	(V^{\alpha_1})^* V^{\alpha_2} \dots V^{\alpha_{2n}} \
	\det{}_{n}G^{\alpha_{2j-1} \alpha_{2k}}(\tau_{2j-1},\tau_{2k}) \\
	\exp\Bigl[\tfrac{\Tr \Delta^2}{\pi^2}\sum_{\substack{i,j=1\\i<j}}^{2n} (-1)^{i+j} \, 
		\ln\bigl(i \xi_0 (\tau_j-\tau_i) \bigr)\Bigr] \\
	\e^{-i (E_d+\Delta E_1) \sum_{j=1}^n (\tau_{2j}-\tau_{2j-1})}.
\end{multline}
The latter two equations represent the central result of this article. Since they are by far too complex
for a direct interpretation, we discuss here two limiting cases. The first one consists in the scalar
problem with only one channel $\alpha$. In the second one we suppress the off-diagonal terms,
$\alpha \neq \alpha'$, in the matrix $W^{\alpha \alpha'}$. 

If we assume that the number $N$ of quantum channels $\alpha$ is equal to $1$, the present model can 
be interpreted as an extension to all orders of perturbation of ND's x-ray absorption model \cite{ND69}, 
or as the spin $1/2$ Kondo problem investigated by Yuval and Anderson \cite{YA70} where the occupancy
of the localized state $\phi_d$ corresponds to the directions of the impurity spin.
The propagator at the $n$-th order of perturbation simplifies to
\begin{multline}
	G(\tau,\tau') = \frac{-i \nu}{\beta} \prod_{j=1}^n \left[
		\frac{\tau-\tau_{2j}}{\tau-\tau_{2j-1}} \frac{\tau'-\tau_{2j-1}}{\tau'-\tau_{2j}} 
	\right]^{\frac{\delta}{\pi}} \e^{-i \mu(\tau-\tau')} \\
	\left[
		\ppf{1}{\tau-\tau'} + \pi \tan\vartheta' \, \delta(\tau-\tau')
	\right],
\end{multline}
for $\tau, \tau'$ lying on an interval where $W \neq 0$. The constants $\beta$ and $\tan\vartheta'$ are
given by (see \cite{ND69}, Eq. (42))
\begin{align*}
	\beta 			&= 1 - 2 \pi g \tan\vartheta + \pi^2 g^2 / \cos^2\vartheta,\\
	\tan\vartheta' 	&= \tan\vartheta - \pi g / \cos^2\vartheta,
\end{align*}
and $\delta$ is the phase shift.
Since all $G(\tau_{2j-1},\tau_{2k})$ commute and the exponent $\delta/\pi$ is the same
for all factors in the determinant, the latter can be expressed in terms of Cauchy determinants,
which are defined as (see \cite{Polya71}, Chap. 7, 3.)
\begin{equation}
	\text{(Cauchy)}_n = \det{}_n\left[\frac{1}{\tau_{2j-1}-\tau_{2k}}\right] 
	= \frac{\underset{1 \le j < k \le n}{\prod} (\tau_{2j-1}-\tau_{2k-1}) (\tau_{2k}-\tau_{2j})}%
	       {\underset{j,k=1,\dots,n}{\prod}(\tau_{2j-1} - \tau_{2k})}.
\end{equation}
If $\tau_1 < \dots < \tau_{2n}$, we have
\begin{equation}
	\text{(Cauchy)}_n = (-1)^n \, \exp\biggl( 
		\sum_{\substack{i,j=1\\i<j}}^{2n} (-1)^{i + j} \ln|\tau_j-\tau_i|
	\biggr).
\end{equation}
The determinant in $L(\tau_1,\dots,\tau_{2n},t)$ reads ($\tau_{2n+1}=\tau_{2n}=t$)
\begin{multline*}
	\det{}_{n+1}G(\tau_{2j-1},\tau_{2k}) = \left(\frac{-i\nu}{\beta}\right)^{n+1} \,
	\e^{+i\mu \sum_{j=1}^n(\tau_{2j}-\tau_{2j-1})} \\
	\prod_{j=1}^{n+1} \left[ \prod_{k=1}^{n}
		\frac{\tau_{2j-1}-\tau_{2k}}{\tau_{2j-1}-\tau_{2k-1}} \
		\frac{\tau_{2j}-\tau_{2k-1}}{\tau_{2j}-\tau_{2k}}
	\right]^{\frac{\delta}{\pi}} \,
	\det{}_{n+1} \left[ \frac{1}{\tau_{2j-1}-\tau_{2k}}\right],
\end{multline*}
where all singularities must be replaced by the cutoff $i\xi_0$.
The last factor is a Cauchy determinant in which all factors depending on $t$ cancel out, except
the singular contribution $1/(t-t)$ for which we introduce a cutoff, too. Similarly, the double product
in front of the determinant does not depend on $t$, and is nothing else than 
$[i\xi_0^n/\text{(Cauchy)}_n]^2$. Hence
\begin{multline*}
 	\det{}_{n+1}G(\tau_{2j-1},\tau_{2k}) = \left(\frac{-i\nu}{\beta}\right)^{n+1} \,
	\xi_0^{1+2n} \, \e^{+i \sum_{j=1}^n(\tau_{2j}-\tau_{2j-1})} \, (-1)^n \\
	\exp\Bigl[(1-2\tfrac{\delta}{\pi})\sum_{\substack{i,j=1\\i<j}}^{2n} (-1)^{i+j} \, 
		\ln\bigl(i \xi_0 (\tau_j-\tau_i) \bigr)\Bigr],
\end{multline*}
where we have used the identity $\sum_{1 \le i < j \le 2n} (-1)^{i+j} = -n$. 
If we define $\Omega_m = \{(i,j)| 1 \le i < j \le 2n, (i,j) \neq (2m-1,2m) \}$, we obtain for 
$F_W(t)$ a result similar to that of Yuval and Anderson \cite{YA70},
\begin{multline}
	F_W(t) = \frac{1}{|V|^2} \frac{1}{i\xi_0} \sum_{n=2}^\infty \left(\frac{-i\nu}{\beta}\right)^{n}
	\bigl(i\xi |V|\bigr)^{2n} \underset{t_0 < \tau_1 < \dots < \tau_{2n} < t_1}{\int} 
	\dtau_1 \dots \dtau_{2n} \\
	\sum_{m=1}^n \delta(\tau_{2m-1}-t) \, \delta(t-\tau_{2m}) \,
	\exp\Bigl[ -i (E_d + \Delta E_1 - \mu) \sum_{j=1}^n (\tau_{2j}-\tau_{2j-1})\\
	+ (1-2\tfrac{\delta}{\pi} + \tfrac{\delta^2}{\pi^2}) \sum_{(i,j) \in \Omega_m} (-1)^{i+j} 
	\ln\bigl(i \xi_0 (\tau_j-\tau_i)\bigr) \Bigr].
\end{multline}
The function $F_W(t)$ does no longer express a current but a joint probability amplitude
of the presence of a particle in the localized state $\phi_d$ as well as of a particle in the
conduction band described by the operator $C(t)$. If we pass to imaginary times, $\tau_j \to -i \tau_j$,
the latter equation becomes formally equivalent to a grand canonical correlation function of  
a classical one-dimensional gas with two different types of particles: a type $A$ situated at the 
(space) positions $\tau_{2j-1}$, and a type $B$ at the positions $\tau_{2j}$. All particles are
subject to external potentials, $-(E_d + \Delta_1-\mu) \tau$ for the type $A$ and
$+(E_d + \Delta_1-\mu) \tau$ for the type $B$, as well as to logarithmic two-body interactions
which are repulsive between particles of the same type and attractive otherwise. There is no kinetic
energy. The cutoff $\xi_0$
enters as the (hard-core) volume of the particles. In this context, the positions $\tau_{2m-1}$ and
$\tau_{2m}$ that are fixed to the value $t$ by the delta functions must be regarded as having the
distance $\xi_0$ between each other. The correlation function $F_W(t)$ can therefore be interpreted
as the (unnormalized) average density of a couple $A B$ at the point $t$ (with $A$ to the left of $B$).

We can derive an analogous result for the function $F_V(t)$ which represents now the transition rate
of an electron into the localized state. In the same way as above, we find
\begin{multline}
	F_V(t) = \frac{i}{V} \sum_{n=1}^\infty \left(\frac{-i\nu}{\beta}\right)^{n} \bigl(i \xi_0 |V|\bigr)^{2n}
	\underset{t_0 < \tau_1 < \dots < \tau_{2n} < t_1}{\int} \dtau_1 \dots \dtau_{2n} \\
	\sum_{m=1}^{n} \delta(t-\tau_{2m}) \,
	\exp\Bigl[ -i (E_d + \Delta E_1 - \mu) \sum_{j=1}^n (\tau_{2j}-\tau_{2j-1})\\
	+ (1-2\tfrac{\delta}{\pi} + \tfrac{\delta^2}{\pi^2}) \sum_{(i,j) \in \Omega_m} (-1)^{i+j} 
	\ln\bigl(i \xi_0 (\tau_j-\tau_i)\bigr) \Bigr].
\end{multline}
Analogously, this expresses the average number of particles of the type $B$ at the position $t$.

A investigation of the partition function of the one-dimensional gas was performed by Anderson, Yuval and
Hamann \cite{AYH70} by renormalization group methods, a route we will not pursue here.
For the general case with the number of channels $N \ge 2$, a similar mapping onto a classical gas
is not possible in the same way because the determinants of propagators do not form Cauchy
determinants. Within the scope of the present work a similar, but different mapping has not
been found.

Let us now consider the case in which the left and right electrodes are not coupled via $H_W$.
If we assume further that each electrode contains some number of different non-interacting 
quantum channels, we recover a simplified version of the system considered by Matveev and 
Larkin \cite{Matveev92}.
The latter authors computed the phase shifts for different geometries of the electrodes and determined
the current through the resonant level in the barrier in first nontrivial order in perturbation,
corresponding to $[F^R_V(t)]_{(2)}$ in our notation. 
If there is only one quantum channel in each electrode, we have
\[
	\bigl[ F^R_V(t) \bigr]_{(2)} = (V^R)^* \frac{\nu_R}{\beta_R} \int_{t_0}^t \dtau
	\frac{\e^{-i (E_d + \Delta E_1 - \mu_R) (t-\tau)}}{t-\tau}
	\bigl(i\xi_0(t-\tau)\bigr)^{-2\frac{\delta_R}{\pi}+\frac{\delta_R^2}{\pi^2}+\frac{\delta_L^2}{\pi^2}},
\]
with a similar definition of $\beta_R$ as for $\beta$ above. We see that the influence of the left electrode
enters through the closed loop spreading only.
In the limit $t_0 \to -\infty$, this expression becomes time-independent (it is a Gamma function),
whereas scaling arguments reveal the resonant behavior of the current
\[
	\bigl[ F^R_V\bigr]_{(2)} \sim 
	\bigl( E_d + \Delta E_1 - \mu_R \bigr)^{2\frac{\delta_R}{\pi}-\frac{\delta_R^2}{\pi^2}
	                                                           -\frac{\delta_L^2}{\pi^2}}.
\]
Following \cite{Matveev92} the sign of the exponent depends on whether the impurity state
is localized near the left or near the right electrode.

%-----------------------------------------------------------------------------------------------------

\section{Conclusions}

In this paper we studied the tunneling of interacting spinless
electrons through an insulating barrier. It was assumed that the 
interaction was restricted to scattering of conduction electrons
with a single trapped particle in a structureless localized state in 
the barrier. Our central result is given in Eqs.~\eqref{eq:disc:F_W}
and \eqref{eq:disc:F_V} in which we express the tunneling current 
as an infinite perturbation series with respect to the hybridization
term of the Hamiltonian. An extension of the method of Nozi\`eres
and De~Dominicis \cite{ND69} allowed an exact computation of each term in this series,
including the effects of transient interaction between the conduction
electrons and the localized state.
The renormalization of the tunneling amplitude due to the additional
charge in the barrier was taken into account.

The calculations were based on two essential assumptions:
a separable potential, and a completely structureless localized
state. Since the uncoupled systems were assumed to be Fermi liquids,
the model does not apply to one-dimensional systems.
With these limitations, extensions of the model may be considered,
such as to include electron-phonon interactions, many traps,
or disorder in the barrier as well as in the electrodes.
Because of the nanometric size of the system, disorder in the electrodes
situates us in the regime of the universal conductance fluctuations
(UCF) which have been intensively investigated in the last two decades
\cite{UCF}. It is believed that further developments of the present 
work will yield a microscopic description of the
stochastic time behavior of these fluctuations.

%-----------------------------------------------------------------------------------------------------

\appendix

%-----------------------------------------------------------------------------------------------------

\section{Solution of the singular integral equation}
\label{sec:int}

In this appendix we solve the system of coupled singular integral equations obtained from the
Dyson equations \eqref{eq:lowest:Singular_Integral} and \eqref{eq:higher:Dyson_eq} for the
one-body matrix propagator $G^{\alpha \alpha'}(\tau,\tau')$ which we denote in this section
as $\varphi^{\alpha \alpha'}(\tau,\tau')$. Generally, these systems of integral equations
can be written in the form
\begin{equation} \label{eq:int_eq:Integral_Equation_general}
A(\tau) \, \varphi(\tau, \tau') - \frac{1}{\pi i} \int_L  \, 
  \frac{B(\tau'') \, \varphi(\tau'',\tau')}{\tau'' - \tau} \, \dtau'' = f(\tau,\tau').
\end{equation}
Here $A, B$ and $f$ are given $N \times N$ matrix functions, $\varphi$ is the unknown propagator
with the entries $\varphi^{\alpha \alpha'}$, and $L$ is a contour in the complex plane with the 
end points $t$ and $t'$ which may take finite values or
lie at infinity (we do not consider the case where $L$ is a closed contour, i.e. $t = t'$). 
The positive direction on $L$ is chosen to run from $t$ to $t'$.

The technique to solve such integral equations was developed by Muskhelishvili \cite{Muskhelishvili53}
and Vekua \cite{Vekua67} and consists normally in a transformation to a system of ordinary coupled
Fredholm equations. However, the particular form of the coefficients $A$ and $B$ as given in 
Eqs.~\eqref{eq:lowest:A} and \eqref{eq:lowest:B} and their generalization to higher orders
in perturbation allows us to develop a compact, closed solution by a method similar to that
used by Yuval and Anderson \cite{YA70} for the scalar problem. This method is in fact nothing
else than Vekua's technique to regularize discontinuities in the coefficients $A$ and $B$, and
we formulate it for arbitrary smooth contours $L$ in the complex plane.
In Sec.~\ref{sec:int:fund_sol} this solution will be carried out in detail. Furthermore, we will show
its limitation in the solution of more general problems.

% ----------------------------------------------------------------------------------------------------------

\subsection{Transformation to a Hilbert problem}
\label{sec:int:trans_hilbert}

The first step in the solution of Eq. \eqref{eq:int_eq:Integral_Equation_general} consists in a transformation
of the integral equation to a Hilbert boundary value problem. We follow the lines of 
Muskhelishvili and Vekua (see e.g. \cite{Vekua67} \S16), which we expose briefly in the sequel.
 
From now on it is assumed that the functions $A(\tau)$ and $B(\tau)$ are $N \times N$ matrices,
which are - for the theory to be valid - required to be such that
\begin{equation*}
 	S(\tau) \equiv A(\tau) + B(\tau), \qquad
	D(\tau) \equiv A(\tau) - B(\tau),
\end{equation*}
are regular, i.e. $\det S(\tau) \neq 0$ and $\det D(\tau) \neq 0$ for all $\tau \in L$.
Furthermore we assume that all entries of $A(\tau)$ and $B(\tau)$ satisfy a H\"older 
condition on the line $L$ except at some points $\tau_1, \dots, \tau_d \in L$ where they have
finite discontinuities, but where they still are assumed to be such that $S(\tau)$ and $D(\tau)$
remain regular when considering the left and right limits $\tau \to \tau_k \pm 0$ 
($k = 1, \dots, d$) on $L$.

The Hilbert problem is formulated by means of the function $\Psi(z)$, which is holomorphic in the whole
plane, except on the line $L$ where it is continuous from the left and the right side with respect
to the positive direction of $L$ (except possibly at the points $\tau_k$),
\[
	\Psi(z) = \frac{1}{\pi i} \int_L \frac{B(\tau'') \, \varphi(\tau'')}{\tau'' - z} \, \dtau''.
\]
In the terminology of Muskhelishvili and Vekua, such a function is called a \emph{sectionally
holomorphic function}.

The connection between the functions $\Psi$ and $\varphi$ is given by the Plemelj formulae
(see e.g. \cite{Muskhelishvili53}, \S 17 or \cite{Vekua67}, \S 3)
\begin{equation} \label{eq:int_eq:plemelj}
\begin{split}
	\frac{1}{2} \big[ \Psi^+(\tau) - \Psi^-(\tau) \big] &= B(\tau) \, \varphi(\tau,\tau'),\\
	\frac{1}{2} \big[ \Psi^+(\tau) + \Psi^-(\tau) \big] &= \frac{1}{\pi i} \int_L 
		\frac{B(\tau'') \, \varphi(\tau'',\tau')}{\tau'' - \tau} \, \dtau''
\end{split}
\end{equation}
for $\tau \in L$, not a point of discontinuity $\tau_k$. The functions $\Psi^+(\tau)$ and $\Psi^-(\tau)$ 
denote respectively the limiting values of $\Psi(z)$ when $z$ approaches the point $\tau \in L$ from the left or the
right side of $L$ with respect to the positive direction on $L$.
By replacing the integral in equation \eqref{eq:int_eq:Integral_Equation_general} by the second Plemelj
formula \eqref{eq:int_eq:plemelj} we obtain
$$A(\tau) \, \varphi(\tau) = \frac{1}{2} \big[ \Psi^+(\tau) + \Psi^-(\tau) \big] + f(\tau).$$ 
Adding and subtracting the first Plemelj formula \eqref{eq:int_eq:plemelj} from this last expression
leads to
\begin{equation} \label{eq:int_eq:phi_in_Psi}
\begin{split}
	\varphi(\tau) &= [S(\tau)]^{-1} \, \Psi^+(\tau) + [S(\tau)]^{-1} \, f(\tau),\\
	\varphi(\tau) &= [D(\tau)]^{-1} \, \Psi^-(\tau) + [D(\tau)]^{-1} \, f(\tau).
\end{split}
\end{equation}
From these two expressions we obtain the following Hilbert boundary value problem, equivalent to the integral equation 
\eqref{eq:int_eq:Integral_Equation_general},
\begin{equation} \label{eq:int_eq:Hilbert}
	\Psi^+(\tau) = \mathfrak{S}(\tau) \, \Psi^-(\tau) + b(\tau)
\end{equation}
where we have defined
\begin{align}
	\mathfrak{S}(\tau) 	&= S(\tau) \, [D(\tau)]^{-1}   
				 = [ A(\tau) + B(\tau) ] \, [ A(\tau) - B(\tau) ]^{-1} \label{eq:int_eq:def_G}, \\
	b(\tau)		&= \big( \mathfrak{S}(\tau) - \openone \big) \, f(\tau,\tau').
\end{align}	
The matrix $\mathfrak{S}$ is called the \emph{coefficient} of the Hilbert problem.
Once we have found the solution of the Hilbert problem
\eqref{eq:int_eq:Hilbert} the function $\varphi(\tau,\tau')$ can be recovered by any of the 
equations \eqref{eq:int_eq:phi_in_Psi}.

As will be shown below, this matrix $\mathfrak{S}$ is equal to the inverse scattering matrix at the Fermi
levels, and is therefore unitary.
% ----------------------------------------------------------------------------------------------------------

\subsection{Formal solution of the integral equation}
\label{sec:int:formal_sol}

The Hilbert problem \eqref{eq:int_eq:Hilbert} is solved by means of a particular solution
(the so-called \emph{fundamental} solution) of the homogeneous Hilbert problem obtained by putting
$b(\tau) \equiv 0$ in \eqref{eq:int_eq:Hilbert}. This fundamental solution will be computed in 
Sec.~\ref{sec:int:fund_sol} and is given by a
$N \times N$ matrix function $X(z)$, where $z$ is a point in the plane, 
not on $L$. The limits of $X(z)$
on $L$ from the left and from the right with respect to the positive direction on $L$ 
will be denoted by $X^+(\tau)$ and $X^-(\tau)$, respectively, for 
$\tau \in L\backslash\{\tau_1, \dots, \tau_d\}$.

Since $X(z)$ is a solution of the homogeneous Hilbert problem corresponding to \eqref{eq:int_eq:Hilbert},
we have
\[
	\mathfrak{S}(\tau) = X^+(\tau) \, [X^-(\tau)]^{-1}.
\]
By Vekua \cite{Vekua67}, \S 6, the general solution of the 
non homogeneous problem \eqref{eq:int_eq:Hilbert}, having at most polynomial divergence at infinity, 
is then given by the expression 
\[
	\Psi(z) = \frac{X(z)}{2 \pi i} \int_L \frac{[X{}^+(\tau'')]^{-1} \, b(\tau'')}{\tau''-z} \, \dtau''
	+ X(z) \, p(z)
\]
with $p(z)$ a matrix of size $N \times N$ with arbitrary polynomials as entries which 
must be determined by the boundary conditions. In the physical applications
to follow, it is sufficient to notice that the function $\varphi(\tau,\tau')$
is a Green's function which must vanish as its arguments go to infinity. 
Since the function $X(z)$ remains bounded at infinity (as we will see later) these polynomials
must identically vanish, which allows us to drop the term $X(z) \, p(z)$ in the sequel.

\begin{figure}[t]
\begin{center}
	\input{figure4.tex}
	\caption{Contour $C$ for the integration of formula \eqref{eq:int_eq:remaining_integral}.}
	\label{fig:cont_C}
\end{center}
\end{figure}
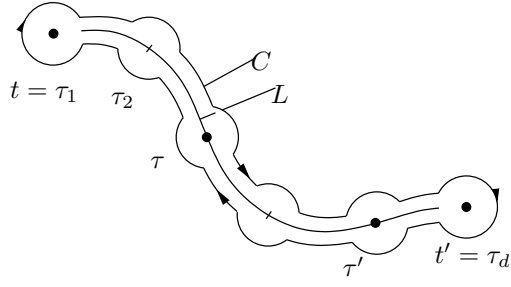

The boundary values $\Psi^+(\tau)$ and $\Psi^-(\tau)$ follow again from the Plemelj formulae
and are given by
\begin{align*}
	\Psi^+(\tau) &= X^+(\tau) \, \Big\{ + \frac{1}{2} [X^+(\tau)]^{-1} \, b(\tau) +
		\frac{1}{2 \pi i} \int_L \frac{[X^+(\tau'')]^{-1} \, b(\tau'')}{\tau'' - \tau} \, \dtau'' 
 		\Big\},\\
	\Psi^-(\tau) &= X^-(\tau) \, \Big\{ - \frac{1}{2} [X^+(\tau)]^{-1} \, b(\tau) +
		\frac{1}{2 \pi i} \int_L \frac{[X^+(\tau'')]^{-1} \, b(\tau'')}{\tau'' - \tau} \, \dtau'' 
		\Big\}.\\
\end{align*}
The solution of the original integral equation \eqref{eq:int_eq:Integral_Equation_general} can then be computed
by any of the equations \eqref{eq:int_eq:phi_in_Psi} and takes the form
\begin{equation} \label{eq:int_eq:solution_Integral_1}
	\varphi(\tau,\tau') = \mathcal{A}(\tau) \, f(\tau) - \frac{1}{\pi i} \, Z(\tau)
	 	\int_L \frac{\mathcal{B}(\tau'') \, f(\tau'')}{\tau'' - \tau} \, \dtau''
\end{equation}		
where we have defined
\begin{align}
	\mathcal{A}(\tau) &= \frac{1}{2} \ \Big\{ [S(\tau)]^{-1}   + [D(\tau)]^{-1}    \Big\}, \label{eq:int_eq:mathcal_A} \\
	\mathcal{B}(\tau) &= \frac{1}{2} \ \Big\{ [X^+(\tau)]^{-1} - [X^-(\tau)]^{-1}  \Big\}, \label{eq:int_eq:mathcal_B} \\
	Z(\tau)   &= [S(\tau)]^{-1} \ X^+(\tau) = [D(\tau)]^{-1} \, X^-(\tau). \label{eq:int_eq:Z} 
\end{align}	
The fact that the inhomogeneous term $f(\tau,\tau')$ is proportional to the free propagator and that
$\mathcal{B}(\tau)$ is the difference between the limiting values of the inverse fundamental solution
allows us to evaluation the remaining integral in \eqref{eq:int_eq:solution_Integral_1},
\begin{equation} \label{eq:int_eq:remaining_integral}
	I = \int_L \frac{\mathcal{B}(\tau'') \, f(\tau'')}{\tau'' - \tau} \, \dtau''.
\end{equation}
Let us write 
\begin{equation} \label{eq:int_eq:def_h(z)}
	f(\tau) = a \, \ppf{1}{\tau - \tau'} + b \, \delta(\tau-\tau')
\end{equation}
where $a$ and $b$ are $N \times N$ matrices that may depend on $\tau'$ but not on $\tau$.
Such $f$ can be continued analytically to the remaining
complex plane, and vanish at $|z| \to \infty$.
Considering the term proportional to $a$ only and assuming that $\tau \neq \tau'$,
the integral \eqref{eq:int_eq:remaining_integral}  can be calculated 
explicitly by the integration of $[X(z)]^{-1} a / \big((z-\tau)(z-\tau')\big)$ over 
the contour $C$ indicated in Fig. \ref{fig:cont_C}. 
In fact, by the theorem of residues, we have on the one hand
\[
	\oint_C \frac{[X(z)]^{-1}  \, a \, \dz}{(z-\tau)(z-\tau')} = 0.
\]	
Since $X(z)$ is the fundamental solution, we will see in Sec.~\ref{sec:int:fund_sol} that its entries are 
non-vanishing, analytic functions everywhere in $\mathbb{C}$ except on $L\backslash \{\tau_1,\dots,\tau_d\}$ 
where it is continuous from the left and the right. At the points $\tau_1, \dots, \tau_d$ 
it either has integrable poles or vanishes.  
Moreover, we will see that $X(z)$ remains bounded at infinity, and hence there is no residue at
infinity.
On the other hand, the integral on the contour $C$ can be split as follows
\begin{multline*}
	\oint_C \frac{[X(z)]^{-1} \, a \, \dz}{(z-\tau)(z-\tau')} = 
		\int_L \frac{[X^+(\tau'')]^{-1}  \, a \,  \dtau''}{(\tau''-\tau)(\tau''-\tau')}
	  - \int_L \frac{[X^-(\tau'')]^{-1}  \, a \,  \dtau''}{(\tau''-\tau)(\tau''-\tau')}\\
	  - \pi i \frac{[X^+(\tau)]^{-1}  + [X^-(\tau)]^{-1} }{\tau-\tau'} \, a
	  - \pi i \frac{[X^+(\tau')]^{-1} + [X^-(\tau')]^{-1}}{\tau'-\tau} \, a.
\end{multline*}
Notice that the singularities at the points of discontinuity (if there are any) are of order 
$z^{-\eps}$  for certain $0 < \eps < 1$ and do not contribute.
When combining the last two equations, we obtain the integral,
\begin{multline*}
	I =  \frac{\pi i}{2} \ \ppf{1}{\tau-\tau'} \bigg\{ 
			\Big( [X^+(\tau)]^{-1}  + [X^-(\tau)]^{-1}  \Big)  \, a
		\\ - 
			\Big( [X^+(\tau')]^{-1} + [X^-(\tau')]^{-1} \Big)  \, a \\
				+ \frac{1}{\pi i} \Big( [X^+(\tau')]^{-1} - [X^-(\tau')]^{-1} \Big) \, b
		\bigg\}.
\end{multline*}
where we have now added the contribution arising from the integration over the delta function.
Since we assumed explicitly that $\tau \neq \tau'$ the last expression has to be considered as 
being a Cauchy principal value distribution. The singular contribution for $\tau = \tau'$ can be 
computed from the Poincar\'e-Bertrand formula
\[
\begin{split}
	\ppf{1}{\tau''-\tau} \ppf{1}{\tau''-\tau'} = \
	&\ppf{1}{\tau-\tau'} \left\{
		\ppf{1}{\tau''-\tau} - \ppf{1}{\tau''-\tau'} \right\} \\[8pt]
	&+ \pi^2 \delta(\tau''-\tau) \delta(\tau''-\tau')
\end{split}
\]
and is given by
\[
	\frac{\pi^2}{2} \Big( [X^+(\tau)]^{-1} - [X^-(\tau)]^{-1} \Big) \, a \ \delta(\tau-\tau').
\]
These results are introduced into the expression \eqref{eq:int_eq:solution_Integral_1} which leads to
the solution of the integral equation \eqref{eq:int_eq:Integral_Equation_general},
\begin{equation} \label{eq:int_eq:solution_Integral_2}
\begin{split}
	\varphi(\tau,\tau') = \  &Z(\tau) \, [Z(\tau')]^{-1} \, \left( 
		\mathcal{A}(\tau') \, a + \frac{1}{\pi i} \, \mathcal{C}(\tau') \, b \right) \
		\ppf{1}{\tau-\tau'}\\
		 + \ &\left( \mathcal{A}(\tau') \, b + \pi i \, \mathcal{C}(\tau') \, a \right) \
		 \delta(\tau-\tau')
\end{split}
\end{equation}
where $\mathcal{A}(\tau)$ and $Z(\tau)$ are defined by \eqref{eq:int_eq:mathcal_A} and \eqref{eq:int_eq:Z},
where
\begin{equation} \label{eq:int_eq:mathcal_C}
	\mathcal{C}(\tau) = \frac{1}{2} \left\{ \bigl[S(\tau)\bigr]^{-1} - \bigl[D(\tau)\bigr]^{-1} \right\}.
\end{equation}

% ----------------------------------------------------------------------------------------------------------

\subsection{Fundamental solution of the homogeneous Hilbert problem obtained from the integral equation}
\label{sec:int:fund_sol}

In this section we determine the fundamental solution of the homogeneous Hilbert problem
\begin{equation} \label{eq:int_eq:hom_Hilbert}
	\Phi^+(\tau) = \mathfrak{S}(\tau) \, \Phi^-(\tau)
\end{equation}
for $\tau \in L\backslash\{\tau_1, \dots, \tau_d\}$. The matrix
$\mathfrak{S}(\tau)$ is defined by Eq.~\eqref{eq:int_eq:def_G} for the different cases imposed by the
underlying physical problems.

%...........................................................................................................

\subsubsection{Characterization of the fundamental matrix}
\label{sec:int:fund_matrix}

Following Vekua \cite{Vekua67}, \S 5, a fundamental solution of \eqref{eq:int_eq:hom_Hilbert} is a matrix, 
$X(z)$, characterized by the following two properties: 
\begin{enumerate}[\quad (a)]
	\item \label{enum:int_eq:criterium1} If $X(z)$ is a fundamental matrix, then its determinant, $\det X(z)$,
	does not vanish anywhere in the finite part of the complex plane.
	\item \label{enum:int_eq:criterium2} Let $\overset{\beta}{\zeta}(z)$ be the vector formed from the column 
	$\beta$ ($\beta = 1, \dots, N$) of $X(z)$,
	$\overset{\beta}{\zeta}_\alpha(z) = X_{\alpha \beta}(z)$ for $\alpha = 1, \dots, N$, and
	$(-\kappa_\beta)$ its degree at infinity 
	(i.e. $\overset{\beta}{\zeta}_\alpha(z) = \mathcal{O}(z^{-\kappa_\beta})$ for $z \to \infty$). 
	Then the determinant $\det Y(z)$ of the matrix 
	$Y_{\alpha \beta}(z) = z^{\kappa_\beta} \overset{\beta}{\zeta}_\alpha(z)$
	has a finite non-zero value at infinity.
\end{enumerate}

Any system of solutions (represented by such a matrix $X$) 
satisfying these two properties will be called fundamental, and
it can be shown (\cite{Vekua67}, \S 5) that the set of degrees at infinity $\{\kappa_1, \dots, \kappa_N\}$
is the same for all fundamental systems.

%...........................................................................................................

\subsubsection{Regularization of the discontinuities}
\label{sec:int:regularization}

Consider the homogeneous Hilbert problem \eqref{eq:int_eq:hom_Hilbert}. We assume first that
the matrix $\mathfrak{S}(\tau)$ is a constant, unitary matrix $\mathfrak{S}$ on the whole line $L$,
$\mathfrak{S}(\tau) = \mathfrak{S}, \forall \tau \in L$. 

We choose an auxiliary line $L'$ in the complex plane such that $\mathcal{C} = L \cup L'$ forms
a simple, closed, smooth contour with the positive direction running from $t$ to $t'$ on $L$ and
from $t'$ to $t$ on $L'$. The domain bounded by $\mathcal{C}$, denoted by $\mathcal{D}^+$, is 
assumed to lie to the left of $\mathcal{C}$ with respect to the positive direction. The unbounded 
region outside the contour is denoted by $\mathcal{D}^-$. As long as these requirements are
fulfilled, the choice of $L'$ is arbitrary and has no influence on the result.
The Hilbert problem \eqref{eq:int_eq:hom_Hilbert} can be extended to the contour $\mathcal{C}$
by putting $\mathfrak{S}(\tau) = \openone$, the $N \times N$ unit matrix, on $L'$. With this
choice $\mathfrak{S}(\tau)$ has discontinuities at the points $t$ and $t'$.

The method presented below consist in a simplified version of Vekua's technique to suppress
the discontinuities of $\mathfrak{S}(\tau)$. Generally, this transforms the coefficient
$\mathfrak{S}(\tau)$ into a smooth function only and does not give a solution to
\eqref{eq:int_eq:hom_Hilbert}. However, in the present case of a constant, unitary coefficient
it is already sufficient to solve the problem.

We define the matrices
\begin{equation} \label{eq:int_eq:def_gamma}
\begin{split}
	\gamma &= \big[\mathfrak{S}(t+0)\big]^{-1} \, \mathfrak{S}(t-0) = \mathfrak{S}^{-1},\\
	\gamma' &= \big[\mathfrak{S}(t'+0)\big]^{-1} \, \mathfrak{S}(t'-0) = \mathfrak{S}.
\end{split}
\end{equation}	
The unitarity of $\mathfrak{S}$ assures that we can find a unitary matrix $B_0$ that
diagonalizes simultaneously $\gamma$ and $\gamma'$ such that
\begin{equation} \label{eq:int_eq:def_lambda}
\begin{split}
	\lambda  &= B_0{\!\!}^{-1} \, \gamma \, B_0,\\
	\lambda' &= B_0{\!\!}^{-1} \, \gamma' \, B_0,
\end{split}
\end{equation}
are diagonal and inverse to each other.
Further we may choose the real numbers $\delta_\alpha$, with 
$-\pi < \delta_\alpha < +\pi, \alpha = 1, \dots, N,$ 
such that
\begin{equation}
\begin{split}
	\lambda_\alpha  &= \e^{-2 i \delta_\alpha},\\
	\lambda'_\alpha &= \e^{+2 i \delta_\alpha}.
\end{split}
\end{equation}
The condition $-\pi < \delta_\alpha < +\pi$ is a mathematical requirement that ensures the convergence
of the expressions, but it is shown in Appendix~\ref{sec:phase_shifts} that the same restrictions result
from purely physical arguments. Even more, the physical constraints fix uniquely the angles $\delta_\alpha$
to be obtained from the phase shifts from the quotient of the scattering matrices immediately before
and after a transition (see Eq.~\eqref{eq:int_eq:def_gamma}).

We choose an arbitrary point $z_0$ in $\mathcal{D}^+$ and define the diagonal matrix functions
$\xi_0, \xi_1$ and $\xi$ as
\begin{align}
	\bigl(\xi_0(z)\bigr)_\alpha &= (z-z_0)^{\frac{\delta_\alpha}{\pi}},	\label{eq:int_eq:xi_0}\\
	\bigl(\xi_1(z)\bigr)_\alpha &= (z-t)^{\frac{\delta_\alpha}{\pi}},	\label{eq:int_eq:xi_1}\\
	\bigl(\xi(z)\bigr)_\alpha   &= \left(\frac{z-t}{z-z_0}\right)^{\frac{\delta_\alpha}{\pi}}, \label{eq:int_eq:xi}
\end{align}
for $\alpha = 1, \dots, N$. Let $\ell$ be a not self-intersecting smooth line that starts at 
$z_0$, intersects $\mathcal{C}$ once in $t$ and goes then to infinity.
The whole line $\ell$ is chosen to be the branch line of $\xi_0(z)$, its part starting at $t$
the branch line of $\xi_1(z)$, and its segment $(z_0,t)$ the branch line of $\xi(z)$. 
Similarly we construct the 
functions $\xi'_0(z), \xi'_1(z)$ and $\xi'(z)$ which are given by Eqs.~\eqref{eq:int_eq:xi_0}--\eqref{eq:int_eq:xi}
with the substitutions $t'$ for $t$, and with a branch line $\ell'$ that starts at $z_0$, crosses 
$\mathcal{C}$ at $t'$ and goes then to infinity.

In this way, the functions $\xi(z)$ and $\xi'(z)$ are holomorphic in $\mathcal{D}^-$, whereas
$\xi_1(z)$ and $\xi'_1(z)$ are holomorphic in $\mathcal{D}^+$. All four functions can be continued
analytically to the contour $\mathcal{C}$ with the exception of their respective branch points $t$ or $t'$.

Now introduce the matrices
\begin{equation} \label{eq:int_eq:def_A}
\begin{split}
	A_0 &= 
	\mathfrak{S}(t + 0) \, B_0 \, \bigl[\xi_0(t + 0)\bigr]^{-1} = 
	\mathfrak{S}(t - 0) \, B_0 \, \bigl[\xi_0(t - 0)\bigr]^{-1}, \\
	A'_0 &= 
	\mathfrak{S}(t' + 0) \, B_0 \, \bigl[\xi_0(t' + 0)\bigr]^{-1} = 
	\mathfrak{S}(t' - 0) \, B_0 \, \bigl[\xi_0(t' - 0)\bigr]^{-1}, \\
\end{split}
\end{equation}	
and define the functions
\begin{align*}
	Q(z)  &= A_0   \, \xi_1(z),\\
	Q'(z) &= \bigl[Q(t')\bigr]^{-1} \, A'_0 \, \xi'_1(z),
\end{align*}
for $z \in \mathcal{D}^+$, and 
\begin{align*}
	P(z)  &= B_0 \, \xi(z),\\
	P'(z) &= \bigl[P(t')\bigr]^{-1} \, B_0 \, \xi'(z),
\end{align*}
for $z \in \mathcal{D}^-$.
These functions allow us to substitute the sectionally holomorphic function $\Psi(z)$ in 
the homogeneous Hilbert problem \eqref{eq:int_eq:hom_Hilbert} by the new functions
\begin{equation} \label{eq:int_eq:def_Psi_tilde}
 	\widetilde{\Psi}(z) = 
	\begin{cases}
		\bigl[ Q'(z) \bigr]^{-1} \, \bigl[ Q(z) \bigr]^{-1} \, \Phi(z) 	&\mbox{for $z \in \mathcal{D}^+$},\\
		\bigl[ P'(z) \bigr]^{-1} \, \bigl[ P(z) \bigr]^{-1} \, \Phi(z) 	&\mbox{for $z \in \mathcal{D}^-$},
	\end{cases}
\end{equation}
which leads to the new Hilbert problem
\begin{equation} \label{eq:int_eq:Hilbert_tilde}
	\widetilde{\Psi}^+(\tau) = \widetilde{\mathfrak{S}}(\tau) \, \widetilde{\Psi}^-(\tau)
\end{equation}
with the new coefficient
\begin{equation}
	\widetilde{\mathfrak{S}}(\tau) = \bigl[Q'(\tau) \bigr]^{-1} \, \bigl[Q(\tau) \bigr]^{-1} \, 
									 \mathfrak{S}(\tau)
  		  	  			  			 P(\tau) \, P'(\tau).
\end{equation}
An investigation of the phases of the six functions $\xi_{...}(\tau)$ and $\xi'_{...}(\tau)$
shows that $\widetilde{\mathfrak{S}}(\tau)$ equals the unit matrix on the whole contour $\mathcal{C}$,
including the points $t$ and $t'$. The Hilbert problem \eqref{eq:int_eq:hom_Hilbert} has therefore
been reduced to the trivial problem $\widetilde{\Psi}^+(\tau) \equiv  \widetilde{\Psi}^-(\tau)$,
which consist in seeking a function which is holomorphic in the whole plane $\mathbb{C}$. This condition,
together with the requirements of Sec.~\ref{sec:int:fund_matrix}, shows that the fundamental solution
of \eqref{eq:int_eq:Hilbert_tilde} must be a constant matrix, which we choose to be the unit matrix,
$\widetilde{X}(z) \equiv \openone$. Inverting the relations \eqref{eq:int_eq:def_Psi_tilde} leads to 
the fundamental solution $X(z)$ of the original problem \eqref{eq:int_eq:hom_Hilbert},
\begin{equation}
	X(z) = \begin{cases}
		Q(z) \, Q'(z)	&\text{for $z \in \mathcal{D}^+$},\\
		P(z) \, P'(z)	&\text{for $z \in \mathcal{D}^-$},
	\end{cases}
\end{equation}
and it can readily be verified that this function also satisfies the requirements of 
Sec.~\ref{sec:int:fund_matrix}.

%...........................................................................................................

\subsubsection{Extension to a coefficient fluctuating between two values}

Let us now extend the above method to the case where $\mathfrak{S}(\tau)$ is of the form
\[
\mathfrak{S}(\tau) = 
	\begin{cases}
		\mathfrak{S}		&\text{for $\tau \in (\tau_j, \tau'_j),\  j = 1 \dots n$}\\
		\openone	&\text{otherwise}
	\end{cases}
\]
where $\tau_1, \tau'_1, \tau_2, \tau'_2, \dots, \tau_n, \tau'_n$ are subsequent points (with respect
to the positive direction) on $L$,
and where $\mathfrak{S}$ is a unitary matrix.

We now apply the same regularization method as above. 
Let us introduce the functions $\overset{k}{\xi}_0(z), \overset{k}{\xi}_1(z)$ and $\overset{k}{\xi}(z)$,
as well as $\overset{k}{\xi'}_0(z), \overset{k}{\xi'}_1(z)$ and $\overset{k}{\xi'}(z)$, for $k=1, \dots,n$,
which are defined by Eqs.~\eqref{eq:int_eq:xi_0}--\eqref{eq:int_eq:xi} with the substitutions $\tau_k$ 
for $t$ and $\tau'_k$ for $t'$, respectively, and the appropriate choice of their branch lines.
Consider now the points $\tau_1$ and $\tau'_1$ and define
\begin{align*}
	\overset{1}{P}(z) &= B_0 \, \overset{1}{\xi}(z),	&	
		\overset{1}{Q}(z) &= \overset{1}{A_0} \, \overset{1}{\xi}_1(z), \\
	\overset{1}{P'}(z) &= \bigl[\overset{1}{P}(\tau'_1)\bigr]^{-1} \, B_0 \, \overset{1}{\xi'}(z),	&	
		\overset{1}{Q'}(z) &= \bigl[\overset{1}{Q}(\tau'_1)\bigr]^{-1} \,
		\overset{1}{A'_0} \, \overset{1}{\xi}_1(z), 
\end{align*}
where
\begin{align*}
	\overset{1}{A_0}  &= \mathfrak{S}(\tau_1 \pm 0)  \, B_0 \, \bigl[\overset{1}{\xi_0}(\tau_1 \pm 0)) \bigr]^{-1},\\
	\overset{1}{A'_0} &= \mathfrak{S}(\tau'_1 \pm 0) \, B_0 \, \bigl[\overset{1}{\xi'_0}(\tau'_1 \pm 0))\bigr]^{-1}.
\end{align*}
The new coefficient of the Hilbert problem is given by
\[
	\mathfrak{S}^{(1)}(\tau) = \bigl[\overset{1}{Q'}(\tau)\bigr]^{-1} \, \bigl[\overset{1}{Q}(\tau)\bigr]^{-1} \
					\mathfrak{S}(\tau) \
					\overset{1}{P}(\tau) \, \overset{1}{P'}(\tau).
\]
As we have seen above, in the interval $(\tau_1, \tau'_1)$ this new coefficient is equal to $\openone$.
But since $\mathfrak{S}(\tau)$ takes only the two values $\openone$ and $\mathfrak{S}$, which are both unitary matrices, 
and since $B_0$ diagonalizes $\mathfrak{S}(\tau)$ on the whole line $L$, the module of $\mathfrak{S}^{(1)}$ is
equal to $\openone$ everywhere on $L$. However, the phases vary between the different segments
on the line, such that 
\[
	\mathfrak{S}^{(1)}(\tau) = \begin{cases}
		\e^{2 i \widetilde{\Delta}}	&\text{for $\tau \not\in (\tau_1, \tau'_1)$ and where $\mathfrak{S}(\tau) = \mathfrak{S}$}\\
		\openone                    &\text{otherwise}
	\end{cases}
\]
for the diagonal real matrix $\widetilde{\Delta} = \mathrm{diag}\left(\delta_1, \dots, \delta_N\right)$.

If the original Hilbert problem is written in the form
\[
	\Psi^+(\tau) = \mathfrak{S}(\tau) \, \Psi^-(\tau)
\]
we have now found an equivalent Hilbert problem expressed by
\[
	\left( \bigl[C^+(\tau)\bigr]^{-1} \, \Psi^+(\tau) \right) =
	\mathfrak{S}^{(1)}(\tau) \, \left( \bigl[ C^-(\tau)\bigr]^{-1} \, \Psi^-(\tau) \right)
\]
where
\[
	C^+(\tau) = \overset{1}{Q}(\tau) \, \overset{1}{Q'}(\tau), \qquad
	C^-(\tau) = \overset{1}{P}(\tau) \, \overset{1}{P'}(\tau).
\]
This is a system of scalar (decoupled) Hilbert problems with discontinuities at the points 
$\tau_2, \tau'_2, \dots, \tau_n, \tau'_n$ which can readily be solved. We can use
the functions $\overset{k}{\xi}_{., 0, 1}(\tau)$ to eliminate the jumps of $\mathfrak{S}^{(1)}$
between the segments.
What remains is again a trivial Hilbert problem, $\widetilde{\Psi}^+ \equiv \widetilde{\Psi}^-$, 
for which a fundamental solution
is given by $\widetilde{X} = \openone$. In fact, we introduce into $\Psi(z)$ for each point of discontinuity $\tau_k$ 
the factor
\begin{align*}
		&\overset{k}{\xi}_1(z) \, \overset{k}{\xi'}_1(z) \, 
		\left[ \overset{k}{\xi}_1(\tau'_k) \, \overset{k}{\xi'_0}(\tau'_k + 0) \right]^{-1} 
		&\text{for $z \in \mathcal{D}^+$,}\\
		&\overset{k}{\xi}(z) \, \overset{k}{\xi'}(z) \, 
		\left[ \overset{k}{\xi}(\tau'_k) \right]^{-1} 
		&\text{for $z \in \mathcal{D}^-$.}
\end{align*}
Notice the factors with $\tau'_k$ as argument. Their effect is to assure the
validity of the solution to the Hilbert problem for any choice of the origin of the phases
of complex numbers
(i.e. the axis with respect to which they are measured).
The final fundamental solution to the Hilbert problem,
denoted by $X(z)$, can then be written 
in the form
\begin{equation} \label{eq:int_eq:Psi(z)_extension}
	X(z) = \begin{cases} \displaystyle
		B_0 \, \prod_{k=1}^n \overset{k}{\xi}_1(z) \, \overset{k}{\xi'}_1(z) \,
		B_0^{-1}
			&\text{for $z \in \mathcal{D}^+$}\\
		\displaystyle			
		B_0 \, \prod_{k=1}^n \overset{k}{\xi}(z) \, \overset{k}{\xi'}(z) \, 
			\overset{k}{\xi_0}(\tau'_k) \, \overset{k}{\xi'_0}(\tau'_k+0) \,
		B_0^{-1}
			&\text{for $z \in \mathcal{D}^-$}
	\end{cases}
\end{equation}
where we have chosen to take $\overset{k}{\xi'_0}(\tau'_k+0)$ into the definition of 
$\Psi^-(z)$ and we have multiplied all expressions by $B^{-1}_0$ from the right (which does not 
change the Hilbert problem, but assures that $X(z)$ tends to the unit matrix $\openone$
as $z$ goes to infinity).

%...........................................................................................................

\subsubsection{Limits of the method}
\label{sec:int:limits}

We show briefly why the above method of regularization does not give the solution for the
case where $\mathfrak{S}(\tau)$ is a more general piecewise constant function, taking the values 
$\mathfrak{S}_1, \mathfrak{S}_2, \dots$ on $L$.

For instance, assume that the coefficient has the form
\[
\mathfrak{S}(\tau) = \begin{cases}
	\mathfrak{S}_1 		&\text{for $\tau \in I_1 \equiv (\tau_1, \tau_2)$}\\
	\mathfrak{S}_2			&\text{for $\tau \in I_2 \equiv (\tau_2, \tau_3)$}\\
	\openone	&\text{otherwise}
\end{cases}
\]
for $\tau_1, \tau_2, \tau_3$ some subsequent points on $L$.

For the Hilbert problem
\begin{equation} \label{eq:int_eq:hilbert_xxx}
	\Psi^+(\tau) = \mathfrak{S}(\tau) \, \Psi^-(\tau)
\end{equation}
the regularization above consists in an \emph{ansatz} in form of a product, 
$\Psi(z) = \psi_1(z) \, \psi_2(z)$ where $\psi_k(z)$ is the solution of
\[
	\psi_k^+(\tau) = \begin{cases}
		\mathfrak{S}_k \, \psi_k^-(\tau)	&\text{if $\tau \in I_k$}\\
		\psi_k^-(\tau)			&\text{otherwise}
	\end{cases}
\]
for $k = 1,2$. If we put this expression into Eq.~\eqref{eq:int_eq:hilbert_xxx} we must have
\[
	\psi_1^+(\tau) \, \psi_2^+(\tau) = \mathfrak{S}(\tau) \, \psi_1^-(\tau) \, \psi_2^-(\tau).
\]
This equation is satisfied for $\tau \in I_1$. But if $\tau \in I_2$, the function
$\psi_2(z)$ must simultaneously satisfy the equations
\begin{align*}
	\psi_2^+(\tau) &= \mathfrak{S}_2 \, \psi_2^-(\tau), \\
	\psi_2^+(\tau) &= \left( \bigl[ \psi_1^+(\tau) \bigr]^{-1} \, \mathfrak{S}_2 \, \psi_1^-(\tau)\right) \, 
		\psi_2^-(\tau).
\end{align*}
In other words, since $\psi_1^+(\tau) = \psi_1^-(\tau) \equiv \psi_1(\tau)$ on $I_2$, it is
required that
\[
	\mathfrak{S}_2 = \bigl[\psi_1(\tau)]^{-1} \, \mathfrak{S}_2 \, \psi_1(\tau)	\qquad \forall \tau \in I_2.
\]
This means that $\mathfrak{S}_2$ and $\psi_1(\tau)$ must commute, which is equivalent to saying (by the construction
of $\psi_1$) that $\mathfrak{S}_2$ and $\mathfrak{S}_1$ must commute.
Generally, there is no reason why they should, and the regularization method sketched above cannot
be used to solve the problem. In such cases, Vekua's complete theory which uses
Fredholm integral equations is required.

If, however, $\mathfrak{S}_1$ and $\mathfrak{S}_2$ (or all matrices $\mathfrak{S}_k$ in an extension to more than two different 
matrices) commute, they can all be diagonalized simultaneously, and the procedure of the previous
section applies. The only modification in the
result \eqref{eq:int_eq:Psi(z)_extension} is that the matrices $\widetilde{\Delta}$ in the 
definition of the functions $\xi$ have
additional indices $k$ indicating their origin from $[\mathfrak{S}_k]^{-1} \, \mathfrak{S}_{k-1}$.

% ----------------------------------------------------------------------------------------------------------

\subsection{Solution of the Dyson equations}
\label{sec:int:solution_Dyson_eq}

We turn now to the solution of the Dyson equations \eqref{eq:lowest:Singular_Integral} and 
\eqref{eq:higher:Dyson_eq}. Consider first Eq.~\eqref{eq:lowest:Singular_Integral} which is almost of the 
form of the singular integral equation \eqref{eq:int_eq:Integral_Equation_general} considered in the
previous section. The only difference in \eqref{eq:lowest:Singular_Integral} is the dependence of 
the coefficients $\widetilde{A}(\tau)$ and $\widetilde{B}(\tau)$ on the phase factors
$\e^{i \mu \tau}$. 
In the transformation to a Hilbert problem this difference is of no significance. Therefore, if 
we follow the lines indicated in Sec.~\ref{sec:int:trans_hilbert} we are lead to the equation
\begin{equation} \label{eq:int_eq:Hilbert_phys}
	\Psi^+(\tau) = \widetilde{\mathfrak{S}}(\tau) \, \Psi^-(\tau) + b(\tau)
\end{equation}
with
\begin{align*}
	\widetilde{\mathfrak{S}}(\tau) &=
		\bigl[\widetilde{A}(\tau) + \widetilde{B}(\tau) \bigr] 
		\bigl[\widetilde{A}(\tau) - \widetilde{B}(\tau) \bigr]^{-1} \\
	&= 	\e^{i \mu \tau} \, 
		\bigl[A(\tau) + B(\tau) \bigr] 
		\bigl[A(\tau) - B(\tau) \bigr]^{-1} \, \e^{-i \mu \tau} \\
	& \equiv \e^{i \mu \tau} \, \mathfrak{S}(\tau) \, \e^{-i \mu \tau},
\end{align*}
and
\[
	b(\tau) = \bigl( \widetilde{\mathfrak{S}}(\tau) - \openone \bigr) \, \widetilde{f}(\tau,\tau').
\]
The coefficient $\mathfrak{S}(\tau)$ fulfills the requirement of being a constant
unitary matrix function on the line $L = (\tau_1,\tau_2)$ with the value 
(from Eq.~\eqref{eq:lowest:S_matrix}) 
\begin{equation}  \label{eq:int_eq:value_of_S}
	\mathfrak{S} = 
		\bigl( \openone - \pi (\tan\vartheta + i \openone) g \bigr)
		\bigl[ \openone - \pi (\tan\vartheta - i \openone) g \bigr]^{-1}.
\end{equation}
By comparison with the result of Appendix~\ref{sec:unitary}, Eq.~\eqref{eq:unitary:result},
this function is the inverse scattering matrix for electrons at the Fermi surfaces,
and we can define the hermitian matrix $\Delta$ by
\begin{equation}
	\mathfrak{S} = \e^{2 i \Delta}.
\end{equation}
The eigenvalues of $\Delta$ are the phase shifts which we choose, as discussed in 
Appendix~\ref{sec:phase_shifts}, to lie between $-\pi$ and $+\pi$.

Hence, the homogeneous Hilbert problem
\[
	\Phi^+(\tau) = \mathfrak{S}(\tau) \, \Phi^-(\tau) 
\]
obtained from \eqref{eq:int_eq:Hilbert_phys} by the substitution $\Phi(z) = \e^{-i \mu z} \Psi(z)$
and by putting $b(\tau) \equiv 0$, can be solved by the method described above. This leads to the
fundamental matrix $X(z)$, given by Eq.~\eqref{eq:int_eq:Psi(z)_extension}, 
\[
	X(z) = \begin{cases} \displaystyle
		B_0 \, \xi_1(z) \, \xi'_1(z) \,
		B_0^{-1}
			&\text{for $z \in \mathcal{D}^+$,}\\
		\displaystyle			
		B_0 \, \xi(z) \, \xi'(z) \, 
			\xi_0(\tau'_k) \, \xi'_0(\tau'_k+0) \,
		B_0^{-1}
			&\text{for $z \in \mathcal{D}^-$.}
	\end{cases}
\]
Since the line $L$ coincides with the segment $[\tau_1,\tau_2]$ of the real axis, this formula can be 
simplified. If we choose for the logarithm the usual cut on the negative real axis, the function
$X(z)$ can be written in the form
\begin{equation} \label{eq:int_eq:X_for_Dyson}
	X(z) = \left( \frac{z-\tau_2}{z-\tau_1}\right)^{\frac{\Delta}{\pi}} 
	     = B_0 \, \left( \frac{z-\tau_2}{z-\tau_2}\right)^{\frac{\widetilde{\Delta}}{\pi}} \, B_0^{-1}
\end{equation}
for any $z$, not on $[\tau_1,\tau_2]$, with the definition $t^\Delta = \exp( \Delta \log t )$.

To establish the fundamental solution, $\widetilde{X}(z)$, of the original homogeneous problem obtained from
\eqref{eq:int_eq:Hilbert_phys}, it is not sufficient to substitute 
$\widetilde{X}(z) = \e^{i \mu \tau} X(z)$, because this function is singular
at infinity. For $\widetilde{X}(z)$ to be bounded at infinity, we use the fact that $X(z) \to \openone$
as $|z| \to \infty$, and that a solution to the homogeneous Hilbert problem remains a solution when
it is multiplied by any matrix from the right. Therefore, the matrix function
\begin{equation}
	\widetilde{X}(z) = \e^{i \mu z} \, X(z) \, \e^{-i \mu z}
\end{equation}
fulfills all requirements for a fundamental solution to the original homogeneous Hilbert problem.

The inhomogeneous term $\widetilde{f}$ is given by Eq.~\eqref{eq:lowest:f},
\[
	\widetilde{f}(\tau,\tau') 
		= \e^{i \mu \tau} G_0(\tau-\tau')
		= (-i \nu) \left[ \ppf{1}{\tau-\tau'} + \pi \tan\vartheta \, \delta(\tau-\tau') \right] \,
			\e^{i \mu \tau'}
\]			
and is thus of the form of Eq.~\eqref{eq:int_eq:def_h(z)}. The solution of the integral equation
\eqref{eq:lowest:Singular_Integral} is then provided by Eq.~\eqref{eq:int_eq:solution_Integral_2},
which reads explicitly, for $\tau,\tau' \in \mathbb{R}$,
\begin{multline} \label{eq:int_eq:solution_Singular_Integral}
	G(\tau,\tau') = (-i) S^{-1}(\tau) \,
	\e^{-i \Delta(\tau)} \left| \frac{\tau-\tau_2}{\tau-\tau_1}\right|^{\frac{\Delta}{\pi}}
	\, \e^{-i \mu (\tau-\tau')} \,  
	\e^{+i \Delta(\tau')} \left| \frac{\tau'-\tau_1}{\tau'-\tau_2}\right|^{\frac{\Delta}{\pi}}\\
	\left[ \ppf{1}{\tau-\tau'} + \pi \Theta(\tau') \, \delta(\tau-\tau') \right] \,
	M(\tau') \nu,
\end{multline}
where we have used the identities
\begin{align*}
	\openone - i \tan\vartheta &= (S - \openone) B^{-1}, \\
	\openone + i \tan\vartheta &= (\openone - D) B^{-1}, \\
	D^{-1} - S^{-1}            &= 2 S^{-1} B D^{-1},
\end{align*}
and where we have defined
\begin{align} 
	\Theta(\tau') &= \begin{cases}
		i ( S D - A ) B^{-1}	&	\text{if $\tau' \in (\tau_1,\tau_2)$,} \\
		\tan\vartheta			&	\text{otherwise,}		\label{eq:int_eq:Theta(tau)} 
	\end{cases}  \\
	M(\tau') &= \begin{cases}
		B D^{-1} B^{-1}			&	\text{if $\tau' \in (\tau_1,\tau_2)$,} \\
		\openone				&	\text{otherwise,}		\label{eq:int_eq:M(tau)} 
	\end{cases} \\
	S(\tau) &= \begin{cases}
		S						&	\text{if $\tau  \in (\tau_1,\tau_2)$,} \\
		\openone				&	\text{otherwise,}		\label{eq:int_eq:S(tau)} 
	\end{cases} \\
	\Delta(\tau) &= \begin{cases}
		\Delta					&	\text{if $\tau  \in (\tau_1,\tau_2)$,} \\
		0						&	\text{otherwise.}		\label{eq:int_eq:Delta(tau)} 
	\end{cases}
\end{align}
Now consider the Dyson equation \eqref{eq:higher:Dyson_eq} at higher orders in perturbation. 
The new feature is that the potential
$W$ is nonzero only within the time intervals $(\tau_{2j-1},\tau_{2j}), j=1,\dots,n,$ on the real axis.
Hence only on these intervals the matrix $\mathfrak{S}$ takes the value \eqref{eq:int_eq:value_of_S},
everywhere else it is equal to the unit matrix. By Eq.~\eqref{eq:int_eq:X_for_Dyson} 
the fundamental matrix reads
\begin{equation} \label{eq:int_eq:X_for_Dyson_b}
	X(z) = \Bigl( \prod_{j=1}^{n} \frac{z-\tau_{2j}}{z-\tau_{2j-1}}\Bigr)^{\frac{\Delta}{\pi}} 
	     = B_0 \, \Bigl(\prod_{j=1}^{n} \frac{z-\tau_{2j}}{z-\tau_{2j-1}}\Bigr)^{\frac{\widetilde{\Delta}}{\pi}} \, B_0^{-1},
\end{equation}
and this expression must now replace the $X^+$ in the solution \eqref{eq:int_eq:solution_Singular_Integral},
\begin{multline} \label{eq:int_eq:solution_Singular_Integral_b}
	G(\tau,\tau') = (-i) S^{-1}(\tau) 
	\e^{-i \Delta(\tau)}
		\left| \prod_{j=1}^n \frac{\tau-\tau_{2j}}{\tau-\tau_{2j-1}}\right|^{\frac{\Delta}{\pi}}
	\, \e^{-i \mu (\tau-\tau')} \,  
	\e^{+i \Delta(\tau')} 
		\left| \prod_{j=1}^n \frac{\tau'-\tau_{2j-1}}{\tau'-\tau_{2j}}\right|^{\frac{\Delta}{\pi}}\\
	\left[ \ppf{1}{\tau-\tau'} + \pi \Theta(\tau') \, \delta(\tau-\tau') \right] \,
	M(\tau') \nu,
\end{multline}
with corresponding definitions of the matrices $\Theta(\tau), M(\tau), S(\tau)$ and $\Delta(\tau)$ as in
Eqs.~\eqref{eq:int_eq:Theta(tau)}--\eqref{eq:int_eq:Delta(tau)}.

%-----------------------------------------------------------------------------------------------------

\section{The matrix $\mathfrak{S}$ is unitary}
\label{sec:unitary}

We show that the matrix $\mathfrak{S}$ appearing in the Hilbert problem obtained from
the Dyson equations is equal to the scattering matrix $S$
associated to the external potential $W$. 
This is a generalization of a result of Kohn \cite{Kohn51} who determined the phase shift
by similar scattering theory arguments for the scalar case.
In this appendix $H_0$ denotes the free Hamiltonian
and $H = H_0 + W$ the Hamiltonian with the static potential.

%%%%%%%%%%%%%%%%%%%%%%%%%%%%%%%%%%%%%%%%%%%%%%%%%%%%%%%%%%%%%%%%%%%%%%%%%%%%%%%%%%%%%%%%%%%%%%%%%%%%%%%%%%%%

\subsection{$T$ operator and $S$ matrix}
Let $G(z)$ and $G_0(z)$ be the resolvents
\begin{align*}
	G 		&= \big[ z - H \big]^{-1}, \\
	G_0 	&= \big[ z - H_0 \big]^{-1},
\end{align*}
for a point $z$ in the complex plane, not in the spectrum of $H$ or $H_0$.

The operator $T(z)$ is defined as follows
\begin{equation} \label{eq:unitary:def_T_op}
	T(z) = W + W \, G(z) \, W,
\end{equation}
and satisfies the two identities
\begin{equation}
\begin{split} \label{eq:unitary:identities_T_op}
	G_0(z) \, T(z) &= G(z) \, W, \\
	T(z) \, G_0(z) &= W \, G(z).
\end{split}
\end{equation}
Combining the definition \eqref{eq:unitary:def_T_op} with \eqref{eq:unitary:identities_T_op}, the
\emph{Lippmann-Schwinger} equation for the $T$ operator follows
\begin{equation} \label{eq:unitary:T_op_LS}
	T (z) = W + W G_0(z) T(z).
\end{equation}
In a plane wave basis of electrons with momentum $p$ and additional quantum number $\alpha$,
the scattering matrix $S$ is connected to the $T$ operator by the relation
(see \cite{Taylor})
\begin{equation} \label{eq:unitary:relation_S_T_matrix}
	\bra{p' \alpha'} S \ket{p \alpha} = \delta^{(3)}(p-p') \, \delta_{\alpha \alpha'}
	- 2 \pi i \delta(E_{p' \alpha'} - E_{p \alpha}) \, \lim_{\varepsilon \downarrow 0} \, 
	 	\bra{p' \alpha'} T(E_p + i \varepsilon) \ket{p \alpha}.
\end{equation}		

%--------------------------------------------------------------------------------------------------

\subsection{Case of a separable potential}

Consider now the physical situation described in Sec.~\ref{sec:model}. The potential 
$W$ is assumed to be due to a localized scatterer and sufficiently short ranged. Its matrix
elements between the plane wave states $\ket{p \alpha}$ can then be written in the form
(see Eq.~\eqref{eq:model:separable_W})
\[
	W^{\alpha \alpha'}_{p p'} = u^\alpha_p \, W^{\alpha \alpha'} \, u^{\alpha'}_{p'}{}^*
\]
with
\[
	\int \mathrm{d}p \, |u^\alpha_p|^2 = 1 \qquad \text{and} \qquad u^\alpha_{p_\alpha} = 1
\]
for any vector $p_\alpha$ on the ``Fermi surface'' of the channel $\alpha$.
With this envelope function $u^\alpha_p$ we can define the orthonormal localized conduction electron 
state
\[
	\ket{\zeta^\alpha} = \int \mathrm{d}p \, u^\alpha_p \, \ket{p \alpha}
\]
in which the potential decomposes as
\begin{equation} \label{eq:unitary:W_separated}
	W = \sum_{\alpha \alpha'} \ket{\zeta^\alpha} \, W^{\alpha \alpha'} \, \bra{\zeta^{\alpha'}}.
\end{equation}
Taking the matrix elements between $\ket{\zeta^\alpha}$ and $\ket{\zeta^{\alpha'}}$ of the
Lippmann-Schwinger equation \eqref{eq:unitary:T_op_LS} yields
\[
	T^{\alpha \alpha'}(z) = W^{\alpha \alpha'} + \sum_\beta W^{\alpha \beta} \, G_0^\beta(z) \,
	T^{\beta \alpha'}(z),
\]
where $T^{\alpha \alpha'} = \bra{\zeta^\alpha} T \ket{\zeta^{\alpha'}}$, and
\[
	G_0^\beta(z) = \bra{\zeta^\beta} G_0(z) \ket{\zeta^\beta} 
	             = \int \mathrm{d}p \, \frac{|u^\beta_p|^2}{z - E_{p \beta}}.
\]	
In matrix notation for the indices $(\alpha, \alpha')$ this becomes
\[
	T(z) = W + W \, G_0(z) \, T(z)
\]
which has the solution
\begin{equation} \label{eq:unitary:T_matrix}
	T(z) = [\openone - W \, G_0(z)]^{-1} W = W \, [\openone - G_0(z) \, W]^{-1}.
\end{equation}
An expression for the $S$ matrix (seen as a finite matrix with indices $(\alpha,\alpha')$) can be found 
from \eqref{eq:unitary:relation_S_T_matrix} together with \eqref{eq:unitary:T_matrix}. This yields
\begin{align*}
	\bra{p' \alpha'} S \ket{p \alpha} 
	&= \delta^{(3)}(p-p') \, \delta_{\alpha \alpha'} - 2 \pi i \, \delta(E_{p \alpha} - E_{p' \alpha'}) \, 
	\lim_{\varepsilon \downarrow 0} \,
	\bra{p' \alpha'} T(E_{p \alpha'} + i \varepsilon) \ket{p \alpha}\\
	&= \delta^{(3)}(p-p') \, \delta_{\alpha \alpha'} - 2 \pi i \, \delta(E_{p \alpha} - E_{p' \alpha'}) \, \\
	&\qquad
	\sum_{\beta \beta'} \bracket{p' \alpha'}{\zeta^{\beta'}} 
		\left( W [\openone - G_0(E_{p \alpha} + i0) W ]^{-1} \right)^{\beta' \beta}
		\bracket{\zeta^{\beta}}{p \alpha}\\
	&= \delta^{(3)}(p-p') \, \delta_{\alpha \alpha'} - 2 \pi i \, \delta(E_{p \alpha} - E_{p' \alpha'}) \, \\
	&\qquad
	u^{\alpha'}_{p'} \, \left( W [\openone - G_0(E_{p \alpha} + i0) W ]^{-1} \right)^{\alpha' \alpha} \,
	u^\alpha_p{}^*.
\end{align*}
An integration over the momenta $p$ and $p'$ for the fixed energies $E$ and $E'$ then gives	
\begin{align*}
	S^{\alpha' \alpha}(E,E')
		&= \int \mathrm{d}p \, \mathrm{d}p' \, \delta(E'-E_{p'\alpha'}) \delta(E-E_{p \alpha}) \,
			\bra{p' \alpha'} S \ket{p \alpha} \\
		&= \Bigl[ \tilde{\nu}_\alpha \delta_{\alpha \alpha'} - 
			2 \pi i \, \tilde{\nu}_{\alpha'} \, u^{\alpha'}(E) \, 
			\bigl( W [\openone - G_0(E+i0) W]^{-1} \bigr)^{\alpha' \alpha} \,\\
		&\qquad
			u^\alpha(E) \, \tilde{\nu}_{\alpha}(E) \Bigr] \
			\delta(E-E') \\
		&\equiv S^{\alpha' \alpha}(E) \	 \delta(E-E'),
\end{align*}
where $\tilde{\nu}_\alpha(E)$ is the density of states at the total energy $E$ of the system.

We fix the energy of the incoming electrons at the value $\mu_\alpha$ 
and consider a scattering onto an outgoing state with energy $\mu_{\alpha'}$. Since generally
$\mu_\alpha \neq \mu_{\alpha'}$, the energy difference must be contributed by the localized
scatterer, such that 
$E = \mu_\alpha + E_{\text{ls},\alpha} = \mu_{\alpha'} + E_{\text{ls}, \alpha'}$ remains constant,
for $E_{\text{ls},\alpha}$ the initial and $E_{\text{ls},\alpha'}$ the final energy of the 
scatterer. This allows us to calculate
\[
	G_0^\alpha(E+i0) 
	= \int \mathrm{d}p \, \frac{|u^\alpha_{p}|^2}{E - E_{p \alpha} + i0}
	= \int \mathrm{d}p \, \frac{|u^\alpha_{p}|^2}{\mu_{\alpha} - \epsilon_{p \alpha} + i0}
\]		
for $\epsilon_{p \alpha} = E_{p \alpha} - E_{\text{ls},\alpha}$. If 
$\nu_\alpha(\epsilon) = \tilde{\nu}(\epsilon + E_{\text{ls},\alpha})$, and if $\mathcal{P}$
denotes the principal value of the integral, 
\begin{align*}
	G_0^\alpha(E+i0) 
	&= \int \mathrm{d}\epsilon \, 
	   \frac{|u^\alpha_\epsilon|^2 \, \nu_\alpha(\epsilon)}{\mu_\alpha - \epsilon + i0}\\
	&= - \pi i \, \nu_\alpha \, |u^\alpha_{\mu_\alpha}|^2 + \pint \mathrm{d}\epsilon \, 
	\frac{ |u^\alpha_\epsilon|^2 \, \nu_\alpha(\epsilon)}{\mu_\alpha - \epsilon}\\
	&= - \pi i \, \nu_\alpha - \pi \, \nu_\alpha \, \tan\vartheta_\alpha
\end{align*}
where $\nu_\alpha = \nu_\alpha(\mu_\alpha)$. The angle $\vartheta_\alpha$ is given by the relation
(see also \cite{ND69}, Eq.~(39))
\[
	\tan\vartheta_\alpha = \frac{1}{\pi \nu_\alpha} \, \pint \mathrm{d}\epsilon \, 
	\frac{ |u^\alpha_\epsilon|^2 \, \nu_\alpha(\epsilon)}{\epsilon - \mu_\alpha},
\]
and is identical to the angle appearing in Eq.~\eqref{eq:lowest:free_prop}.
Hence,
\[
	S^{\alpha' \alpha}(E) = \nu_\alpha \, \delta_{\alpha \alpha'} - 
		2 \pi i \, \nu_{\alpha'} \, 
		\bigl( W [\openone + \pi(\tan\vartheta + i \openone) \nu W]^{-1} \bigr)^{\alpha' \alpha} \,
		\nu_\alpha,
\]
or, in matrix notation with $\nu = \{ \nu_\alpha \, \delta_{\alpha \alpha'} \}$, 
$\tan\vartheta = \{ \tan\vartheta_\alpha \, \delta_{\alpha \alpha'} \}$, and 
$g = \{ \nu_\alpha W^{\alpha \alpha'} \}$,
\begin{multline} \label{eq:unitary:matrix_S_E}
	S(E) = \bigl(\openone - 2 \pi i \, g \, [ \openone + \pi (\tan\vartheta + i \openone) g]^{-1} \bigr) \nu\\
	     = \bigl(\openone + \pi (\tan\vartheta - i \openone) g \bigr) 
		   \bigl[\openone + \pi (\tan\vartheta + i \openone) g \bigr]^{-1} \nu
\end{multline}

%--------------------------------------------------------------------------------------------------

\subsection{Unitarity of the matrix $\mathfrak{S}$}

In order to check the unitarity of the matrix $S(E)$ given in \eqref{eq:unitary:matrix_S_E}, we investigate
the product $S(E)^* S(E')$:
\[
	\sum_{\alpha''} S^{\alpha \alpha''}(E)^* \, S^{\alpha' \alpha''}(E')
	= \sum_{\alpha''} \int \mathrm{d}p \, \mathrm{d}p' \, \mathrm{d}p'' \,
		\delta(E_{p \alpha} - E) \, \delta(E_{p' \alpha'} - E') \,
		S^{\alpha \alpha''}_{p p''}{}^* S^{\alpha' \alpha''}_{p' p''}.
\]
The unitarity of $S(p,p')$ implies that
\[
	\sum_{\alpha''} \int \mathrm{d}p'' \,		
	S^{\alpha \alpha''}_{p p''}{}^* \, S^{\alpha' \alpha''}_{p' p''} = 
	\delta_{\alpha \alpha'} \, \delta^{(3)}(p-p')
\]
and hence
\begin{multline*}
	\sum_{\alpha''} S^{\alpha \alpha''}(E)^* \, S^{\alpha' \alpha''}(E')\\
	= \int \mathrm{d}p \, \mathrm{d}p' \, \delta(E_{p \alpha} - E) \, \delta(E_{p' \alpha'} - E') \,
		\delta^{(3)}(p-p') 
	= \nu_{\alpha} \, \nu_{\alpha'} \, \delta(E-E').
\end{multline*}
In other words
\[
	S(E)^* \, S(E') = \nu^2 \delta(E-E').
\]
Therefore, the matrix $S(E) \nu^{-1}$, expressing the scattering matrix at fixed energy $E$ per
state at the Fermi surfaces, is unitary. By comparison with Eq.~\eqref{eq:int_eq:value_of_S} we see that
\begin{equation} \label{eq:unitary:result}
	\mathfrak{S}^{-1} \equiv S(E) \nu^{-1} = 
		\bigl( \openone + \pi \,  (\tan\vartheta - i \openone) \, \nu W \bigr) 
		\bigl[ \openone + \pi \,  (\tan\vartheta + i \openone) \, \nu W \bigr]^{-1}
\end{equation}
which is the same matrix that was obtained in the Hilbert problem of Sec.~\ref{sec:lowest} and
\ref{sec:higher} and yields the physical interpretation of this matrix $\mathfrak{S}$.

%-----------------------------------------------------------------------------------------------------

\section{Discussion of the choice of the phase shifts}
\label{sec:phase_shifts}

In Sec.~\ref{sec:int:regularization} we have seen that the numbers $\delta_\alpha$
are not uniquely defined, and can be chosen to be positive or negative within the limits $\pm 1$.
It will now be shown, that the physical constraints determine uniquely all these numbers.

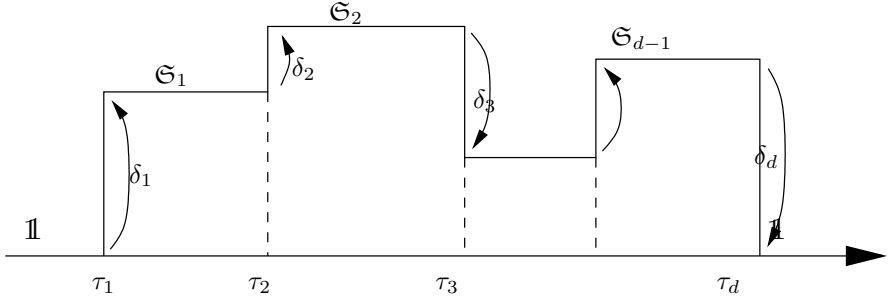
\begin{figure}[t]
\begin{center}
	\input{figure5.tex}
	\caption{A possible form of $\mathfrak{S}(\tau)$.}
	\label{fig:disc:delta_jump}
\end{center}
\end{figure}

For this purpose, let us first focus on the scalar case ($N=1$). The matrix 
$\mathfrak{S}(\tau) = \e^{2 i \delta(\tau)}$ is then a scalar, piecewise constant function,
where $\delta(\tau)$ is a real number. As has been shown in Sec.~\ref{sec:unitary}, 
$\mathfrak{S}(\tau)$ is nothing else than the scattering matrix on the Fermi surface.
This explains why $\mathfrak{S}(\tau)$ is unitary, and that the values $\delta(\tau)$ are the phase 
shifts which the wave functions acquire during the scattering on the local potential $W(\tau)$.
These phase shifts are defined in  $[-\pi, +\pi]$ modulo $2 \pi$ only, and for two different values 
of $\mathfrak{S}(\tau)$ on $L$ the offset from this central interval, $2 \pi m$ for $m$ an integer,
is generally not identical. A possible situation is sketched in Fig.~\ref{fig:disc:delta_jump}.
The arrows in this figure indicate the increments or decrements of $\delta(\tau)$ at the points
of discontinuity. Exactly these in-/decrements enter the exponents of the functions $\xi$ defined
in \eqref{eq:int_eq:xi_0}--\eqref{eq:int_eq:xi}. 
The condition $-\pi < \delta < +\pi$ tells us moreover that
the different phase shifts $\delta(\tau)$ can only differ by less than $\pm 2 \pi$. They must
therefore all have the \emph{same} offset $2 \pi m$ from the interval $[-\pi,+\pi]$.
As long as this condition is satisfied, the value of $m$ is of \emph{irrelevant}. 
For instance we may take $m=0$ (which is perhaps the most intuitive choice for the regions
where $\mathfrak{S} = 1$).

The main condition is therefore seen to be the requirement $-\pi < \delta_\alpha < +\pi$ but
it must be emphasized, that it is so far a purely \emph{mathematical} constraint which ensures
the integrability of the equations. However, there is a \emph{physical} argument for this choice which
was given in ND's paper: only the perturbative solution must be retained. 
Take Eq.~\eqref{eq:int_eq:solution_Singular_Integral_b} for $G(\tau,\tau')$ into consideration.
We know that the function $X^+$ contains the factors
\[
	\left(\frac{\tau-\tau_k}{\tau-\tau'_\alpha}\right)^{\frac{\delta_k}{\pi}}
\]
for the different $\delta_k = \delta(\tau_k + 0) - \delta(\tau_k-0)$ at the points of discontinuity $\tau_k$.
If we let the scattering potentials tend to zero, $W(\tau) \to 0$, we must obtain the free propagator
$G^{(0)}(\tau-\tau')$ given in \eqref{eq:lowest:free_prop}. If in this limit $\delta(\tau_k+0) \to 2 \pi m_+$ and 
$\delta(\tau_k-0) \to 2 \pi m_-$ (for $m_+,m_-$ some integers), there would still remain 
the factor
\[
	\left(\frac{\tau-\tau_k}{\tau-\tau'_k}\right)^{2 ( n_+ - n_- )}
\]
in $G(\tau,\tau')$ which does not appear in $G^{(0)}(\tau-\tau')$. Hence the requirement
$m_+ = m_- \equiv m$ at all $\tau_k$.

In the matrix case where all matrices $\mathfrak{S}(\tau)$ commute
(for any other case, we do not possess a closed solution, as has been 
shown in Sec.~\ref{sec:int:limits}) all matrices can be diagonalized simultaneously. 
The system transforms then into  a set of decoupled scalar problems to which the same arguments 
as above apply.

%-----------------------------------------------------------------------------------------------------

\begin{ack}
We thank B. Altshuler, J.-P. Ansermet, H. Kunz, P.A. Lee, P. Nozi\`eres and T.M. Rice for stimulating
discussions. The support of the Swiss National Fonds is greatfully acknowledged.
\end{ack}

%-----------------------------------------------------------------------------------------------------

\bibliographystyle{unsrt}

%-----------------------------------------------------------------------------------------------------

\end{fmffile}
\end{document}

%% file: figure1.tex
\begin{picture}(0,0)%
\special{psfile=figure1.pstex}%
\end{picture}%
\setlength{\unitlength}{3947sp}%
\begingroup\makeatletter\ifx\SetFigFont\undefined
% extract first six characters in \fmtname
\def\x#1#2#3#4#5#6#7\relax{\def\x{#1#2#3#4#5#6}}%
\expandafter\x\fmtname xxxxxx\relax \def\y{splain}%
\ifx\x\y   % LaTeX or SliTeX?
\gdef\SetFigFont#1#2#3{%
  \ifnum #1<17\tiny\else \ifnum #1<20\small\else
  \ifnum #1<24\normalsize\else \ifnum #1<29\large\else
  \ifnum #1<34\Large\else \ifnum #1<41\LARGE\else
     \huge\fi\fi\fi\fi\fi\fi
  \csname #3\endcsname}%
\else
\gdef\SetFigFont#1#2#3{\begingroup
  \count@#1\relax \ifnum 25<\count@\count@25\fi
  \def\x{\endgroup\@setsize\SetFigFont{#2pt}}%
  \expandafter\x
    \csname \romannumeral\the\count@ pt\expandafter\endcsname
    \csname @\romannumeral\the\count@ pt\endcsname
  \csname #3\endcsname}%
\fi
\fi\endgroup
\begin{picture}(4348,1545)(2190,-2044)
\put(4136,-2044){\makebox(0,0)[lb]{\smash{\SetFigFont{10}{12.0}{rm}localized level}}}
\put(5274,-1170){\makebox(0,0)[lb]{\smash{\SetFigFont{10}{12.0}{rm}$\alpha=R$}}}
\put(2930,-1170){\makebox(0,0)[lb]{\smash{\SetFigFont{10}{12.0}{rm}$\alpha=L$}}}
\put(4058,-693){\makebox(0,0)[lb]{\smash{\SetFigFont{10}{12.0}{rm}insulator}}}
\put(5238,-877){\makebox(0,0)[lb]{\smash{\SetFigFont{11}{13.2}{rm}electrode}}}
\put(2894,-877){\makebox(0,0)[lb]{\smash{\SetFigFont{11}{13.2}{rm}electrode}}}
\end{picture}

%% file: figure4.tex
\begin{picture}(0,0)%
\special{psfile=figure4.pstex}%
\end{picture}%
\setlength{\unitlength}{3947sp}%
\begingroup\makeatletter\ifx\SetFigFont\undefined
% extract first six characters in \fmtname
\def\x#1#2#3#4#5#6#7\relax{\def\x{#1#2#3#4#5#6}}%
\expandafter\x\fmtname xxxxxx\relax \def\y{splain}%
\ifx\x\y   % LaTeX or SliTeX?
\gdef\SetFigFont#1#2#3{%
  \ifnum #1<17\tiny\else \ifnum #1<20\small\else
  \ifnum #1<24\normalsize\else \ifnum #1<29\large\else
  \ifnum #1<34\Large\else \ifnum #1<41\LARGE\else
     \huge\fi\fi\fi\fi\fi\fi
  \csname #3\endcsname}%
\else
\gdef\SetFigFont#1#2#3{\begingroup
  \count@#1\relax \ifnum 25<\count@\count@25\fi
  \def\x{\endgroup\@setsize\SetFigFont{#2pt}}%
  \expandafter\x
    \csname \romannumeral\the\count@ pt\expandafter\endcsname
    \csname @\romannumeral\the\count@ pt\endcsname
  \csname #3\endcsname}%
\fi
\fi\endgroup
\begin{picture}(2926,1636)(1908,-2325)
\put(1908,-1207){\makebox(0,0)[lb]{\smash{\SetFigFont{10}{12.0}{rm}$t=\tau_1$}}}
\put(3996,-2325){\makebox(0,0)[lb]{\smash{\SetFigFont{10}{12.0}{rm}$\tau'$}}}
\put(4573,-2241){\makebox(0,0)[lb]{\smash{\SetFigFont{10}{12.0}{rm}$t'=\tau_d$}}}
\put(3543,-1234){\makebox(0,0)[lb]{\smash{\SetFigFont{10}{12.0}{rm}$L$}}}
\put(3419,-1027){\makebox(0,0)[lb]{\smash{\SetFigFont{10}{12.0}{rm}$C$}}}
\put(2541,-1241){\makebox(0,0)[lb]{\smash{\SetFigFont{10}{12.0}{rm}$\tau_2$}}}
\put(2789,-1642){\makebox(0,0)[lb]{\smash{\SetFigFont{10}{12.0}{rm}$\tau$}}}
\end{picture}

%% file: figure5.tex
\begin{picture}(0,0)%
\special{psfile=figure5.pstex}%
\end{picture}%
\setlength{\unitlength}{3947sp}%
\begingroup\makeatletter\ifx\SetFigFont\undefined
% extract first six characters in \fmtname
\def\x#1#2#3#4#5#6#7\relax{\def\x{#1#2#3#4#5#6}}%
\expandafter\x\fmtname xxxxxx\relax \def\y{splain}%
\ifx\x\y   % LaTeX or SliTeX?
\gdef\SetFigFont#1#2#3{%
  \ifnum #1<17\tiny\else \ifnum #1<20\small\else
  \ifnum #1<24\normalsize\else \ifnum #1<29\large\else
  \ifnum #1<34\Large\else \ifnum #1<41\LARGE\else
     \huge\fi\fi\fi\fi\fi\fi
  \csname #3\endcsname}%
\else
\gdef\SetFigFont#1#2#3{\begingroup
  \count@#1\relax \ifnum 25<\count@\count@25\fi
  \def\x{\endgroup\@setsize\SetFigFont{#2pt}}%
  \expandafter\x
    \csname \romannumeral\the\count@ pt\expandafter\endcsname
    \csname @\romannumeral\the\count@ pt\endcsname
  \csname #3\endcsname}%
\fi
\fi\endgroup
\begin{picture}(5323,1853)(2689,-2716)
\put(7176,-2716){\makebox(0,0)[lb]{\smash{\SetFigFont{10}{12.0}{rm}$\tau_d$}}}
\put(7490,-2402){\makebox(0,0)[lb]{\smash{\SetFigFont{10}{12.0}{rm}$\openone$}}}
\put(4507,-1381){\makebox(0,0)[lb]{\smash{\SetFigFont{10}{12.0}{rm}$\delta_2$}}}
\put(7412,-1931){\makebox(0,0)[lb]{\smash{\SetFigFont{10}{12.0}{rm}$\delta_d$}}}
\put(6509,-1185){\makebox(0,0)[lb]{\smash{\SetFigFont{10}{12.0}{rm}$\mathfrak{S}_{d-1}$}}}
\put(5645,-1578){\makebox(0,0)[lb]{\smash{\SetFigFont{10}{12.0}{rm}$\delta_3$}}}
\put(5410,-2716){\makebox(0,0)[lb]{\smash{\SetFigFont{10}{12.0}{rm}$\tau_3$}}}
\put(4742,-1028){\makebox(0,0)[lb]{\smash{\SetFigFont{10}{12.0}{rm}$\mathfrak{S}_2$}}}
\put(4232,-2716){\makebox(0,0)[lb]{\smash{\SetFigFont{10}{12.0}{rm}$\tau_2$}}}
\put(3251,-2716){\makebox(0,0)[lb]{\smash{\SetFigFont{10}{12.0}{rm}$\tau_1$}}}
\put(2819,-2402){\makebox(0,0)[lb]{\smash{\SetFigFont{10}{12.0}{rm}$\openone$}}}
\put(3486,-2049){\makebox(0,0)[lb]{\smash{\SetFigFont{10}{12.0}{rm}$\delta_1$}}}
\put(3643,-1421){\makebox(0,0)[lb]{\smash{\SetFigFont{10}{12.0}{rm}$\mathfrak{S}_1$}}}
\end{picture}